\documentclass[preprint,12pt]{elsarticle}

\usepackage{amssymb}
\usepackage{amsmath}
\usepackage{hyperref}
\usepackage{multirow}
\usepackage{comment}
\usepackage{subfigure}
\usepackage{bm}

\usepackage[margin=1.5cm]{geometry}
\usepackage{tabularray}
\usepackage[tiny, center, uppercase]{titlesec}
\usepackage{float}
\usepackage{graphicx}
\usepackage{dcolumn}
\usepackage[final,commandnameprefix=ifneeded]{changes}

\journal{Powder Technology journal}

\begin{document}

\begin{frontmatter}




\title{Mixing soft and rigid particles in a hopper: soft particles induce flow intermittency and avalanches}


\author[a]{Saeed Alborzi} 

\affiliation[a]{organization={Department of Mechanical and Industrial Engineering, Northeastern University},
            addressline={360 Huntington Ave.}, 
            city={Boston},
            postcode={02115}, 
            state={MA},
            country={USA}}

\author[a,b,c,*]{Sara M. Hashmi}
\fntext[*]{s.hashmi@northeastern.edu}

\affiliation[b]{organization={Department of Chemical Engineering},
            addressline={360 Huntington Ave.}, 
            city={Boston},
            postcode={02115}, 
            state={MA},
            country={USA}}
\affiliation[c]{organization={Department of Chemistry and Chemical Biology},
            addressline={360 Huntington Ave.}, 
            city={Boston},
            postcode={02115}, 
            state={MA},
            country={USA}}

\begin{abstract}
Instabilities and avalanches in granular flows represent hallmarks of failure: they can both disrupt industrial process flows and signal dangerous conditions, like those in grain silos and snowy mountaintops.  We investigate intermittency and avalanches in the gravity-driven flow of granular materials through a quasi-2D hopper.  Mixtures of rigid polypropylene and soft polyacrylamide particles flow through a hopper constriction.  A combination of high-speed imaging, particle identification and tracking analyses enable us to measure quantities including particle velocities, outflow rates, the time intervals between consecutive particle exits, and the geometric properties of any temporary arches that form during a flow test.  As the fraction of rigid particles increases in the mixture, the velocity of exiting particles increases.  So too, however, does the probability of complete clogging.  While soft particles exhibit slower velocities at the hopper exit, they also facilitate greater overall discharge rates.  Simultaneously, the presence of soft particles induces both intermittency and avalanches in the flow.  In this case, arches that could permanently block the flow are more likely to be temporary in nature.  Interestingly, the identity of the particle that falls first from a temporary arch correlates linearly with the mixing fraction in the overall sample.  That is, soft particles, despite their correlation with flow instabilities, are not significantly more likely to fall first from a temporary arch.  Investigating the arch geometries suggests that the particle forming the largest bond angle with its neighbors is the one that causes an arch to fail, regardless of being soft or rigid.
\end{abstract}


\begin{highlights}

\item Adding soft particles to rigid systems induces flow intermittency and avalanches.
\item Soft particles flow slower, but clog less, allowing faster overall discharge rates.
\item The geometry of temporary arches reveals the origin of subsequent avalanches.


\end{highlights}


\end{frontmatter}



\section{Introduction}
\vspace{1em}
Hopper systems are commonly seen in many industrial applications, including biopharmaceuticals \cite{faqih2010constitutive, salish2024risks, hancock2019wall}, agriculture \cite{landry2005performances,smith1931feeding, gandia2022effect}, food \cite{iqbal2006effect, sousa2021design, juliano2010food}, mining \cite{rojas2019case, wang2020discharge, liu2023segregation}, etc. When dealing with different types of granular materials within these instruments, it is vital to understand the flow characteristics as it impacts process efficiency \cite{bembenek2024mathematical}, risk assessment \cite{mehos2016hopper}, and failure modes \cite{rotter2008silo, ranjan2022failure}. The applications of such assessments range across various length scales, from microns in biological flows \cite{sano2011flow} and active matter \cite{needleman2017active}, to meters in environmental flows \cite{coussot2005rheometry} and pedestrians passage through doors \cite{nicolas2019mechanical}. For instance, the flow of blood cells through narrowed vessels caused by the accumulation of fatty tissues or extracellular matrices on arterial walls can lead to elevated blood pressure and atherosclerosis \cite{zhou2023wall}. In urban sciences, walkability and pedestrian permeability are two important factors for constructing and planning sustainable transport and pedestrian areas in both small towns and major cities \cite{droin2024does}.

The flow of granular materials through hoppers shows interesting features. In rigid systems, clogs occur permanently if the cross-section width shrinks to less than 5 times the particle diameter ($w/d<5$) \cite{mankoc2009role,zuriguel2005jamming}. For deformable particles, such as soft hydrogels \cite{nikoumanesh2023effect, nikoumanesh2024elucidating}, flow continues without permanent stoppages when the outlet size is wider than twice the particle diameter ($w/d\sim2$) \cite{alborzi2022soft, harth2020intermittent}. \added{Among different factors that influence clogging are particle softness \cite{ashour2017silo}, size \cite{zhao2019study}, shape \cite{hafez2021effect}, surface friction, polydispersity \cite{pournin2007influence}, channel geometry \cite{lopez2019effect}, and driving pressure \cite{souzy2022role}.} As reported previously, rigid grains generally drain faster than soft ones, showing no interruptions unless permanently clogged \cite{harth2020intermittent}. In contrast, soft particles flow slower and are less prone to clogging permanently.\deleted{soft} Particles \added{can also} form unstable clogs that disrupt spontaneously, followed by an \textit{Avalanche} \cite{pudasaini2007avalanche}. Avalanches can occur either spontaneously \cite{rajchenbach2005rheology} or due to external forces \cite{swisher2014flow}. In the context of geophysical flows, the study of avalanches, including snow avalanches, is crucial due to their significant natural hazards. When dealing with avalanches, it is important to consider the characteristics of the particles, such as size, shape, deformability, and friction, along with flow conditions like pressure and velocity profiles \cite{pudasaini2007avalanche}.


In rigid systems, intermittency is mostly present when an external force other than gravity is applied. In a study of rigid particle discharge from a gently vibrated vertical silo \cite{mankoc2009role}, clogging probability decreases through two mechanisms when vibration is present: preventing the formation of arches and breaking the formed arches. From a bivariate probability distribution describing the time gaps between particle exits, $\Delta t$, with and without vibrations, it is shown that vibrations mostly encourage the disruption of arches rather than affecting the probability of their formation. Wider arches are less stable against weak perturbations. Additionally, avalanche size $S$, defined as the number of particles that fall between two consecutive arrests, increases when the silo is vibrated. In another study \cite{zuriguel2005jamming}, avalanches were investigated during the discharge of hard grains in a rectangular silo, with clogs disrupted using external pressurized airflow. The avalanche size distribution shows two distinct behaviors: the number of smaller avalanches than a mode increases with increasing $S$, followed by an exponential decay for larger $S$ values. The mean avalanche size $\left<S\right>$ increases with the opening radius until diverging at a critical radius $R_c\simeq$4.94, beyond which clogging never occurs.



Soft systems, in contrast, show a higher tendency to flow intermittently. In \added{a pressure-driven} emulsion flow through a horizontal hopper \cite{hong2022clogging}, time intervals between consecutive exiting droplets, $\Delta t$, are exponentially distributed for fast flow rates, where the flow is continuous and non-intermittent. For slower flow rates, where flow exhibits an avalanche-like regime, $\Delta t$ follows a power-law distribution, with the power-law exponent being 1 at the lowest flow rate and slightly increasing with increasing flow rates. Simulation \added{on 2D particles} suggest that for rigid systems, the behavior shifts from power-law to exponential. Even in highly rigid systems, there is still a very slow flow rate that exhibits power-law behavior, although it is difficult to stabilize experimentally. In a rectangular vertical silo filled with soft hydrogel spheres \cite{harth2020intermittent}, flow intermittency is observed when $w/d\lesssim2$, while rigid airsoft bullets do not show any transient fluctuations under the same conditions. These intermittencies in soft particle flows occur due to particle rearrangements propagating towards the clogged exit. The flow rate of soft hydrogels decreases as the fill height decreases due to the reduction of base pressure, leading to longer flow interruptions. The distribution of time gaps $\Delta t$ between particle exits follows a decaying power-law \deleted{$\sim(\Delta t)^\alpha$ with an exponent $\alpha=-1.85$}. This distribution describes shorter timescales $\Delta t\lesssim3$ $s$ well, while longer timescales $\Delta t\gtrsim 10$ are largely underestimated. The corresponding cumulative distribution of $\Delta t$ also shows a power-law behavior \deleted{$\sim(\Delta t)^\beta$ with an exponent $\beta=-0.85$, following the rule $\beta=\alpha+1$}.



The concept of intermittent flows and avalanches in bidisperse granular systems has been rarely discussed, especially when bidispersity refers to the softness of the particles. In this work, we explore these phenomena in the discharge of mixed soft-rigid granular particles from a quasi-2D hopper. We characterize the flow intermittencies by varying the particle size and mixing fraction of soft and rigid grains.  By also studying uniformly soft and uniformly rigid systems, we can assess the importance of mixing.  We carry out at least 100 trials in each system condition to gather statistics. We define a critical time interval between particle exits, $\Delta t_c$, above which an event is considered to be a pause in the flow, then quantify these instances to determine where they occur most frequently. We quantify the distributions of both these time lapses $\Delta t$ and the subsequent avalanche sizes $S$, which exhibit power-law and exponential tails, respectively, for all mixing fractions. Our results indicate that a power-law distribution governs the intermittent behavior of such bidisperse systems, similar to those of monodisperse ones. Spontaneous avalanches occur more frequently in soft systems compared to rigid ones. Lastly, we address the origins of spontaneous disruptions in particle aggregates, and attribute them to the geometric characteristics of the system. Our results reveal interesting and unexpected phenomena.  In particular, the avalanches that follow temporary arrest events are not well explained by the identity of the first-falling particle nor that of its neighbors.  Rather, the initiation of avalanche events strongly correlate with larger bond angles between the first-falling particle and its neighbors. 


\vspace{1em}
\section{Methods}
\begin{figure}[h]
\centering
\includegraphics[width=0.5\columnwidth]{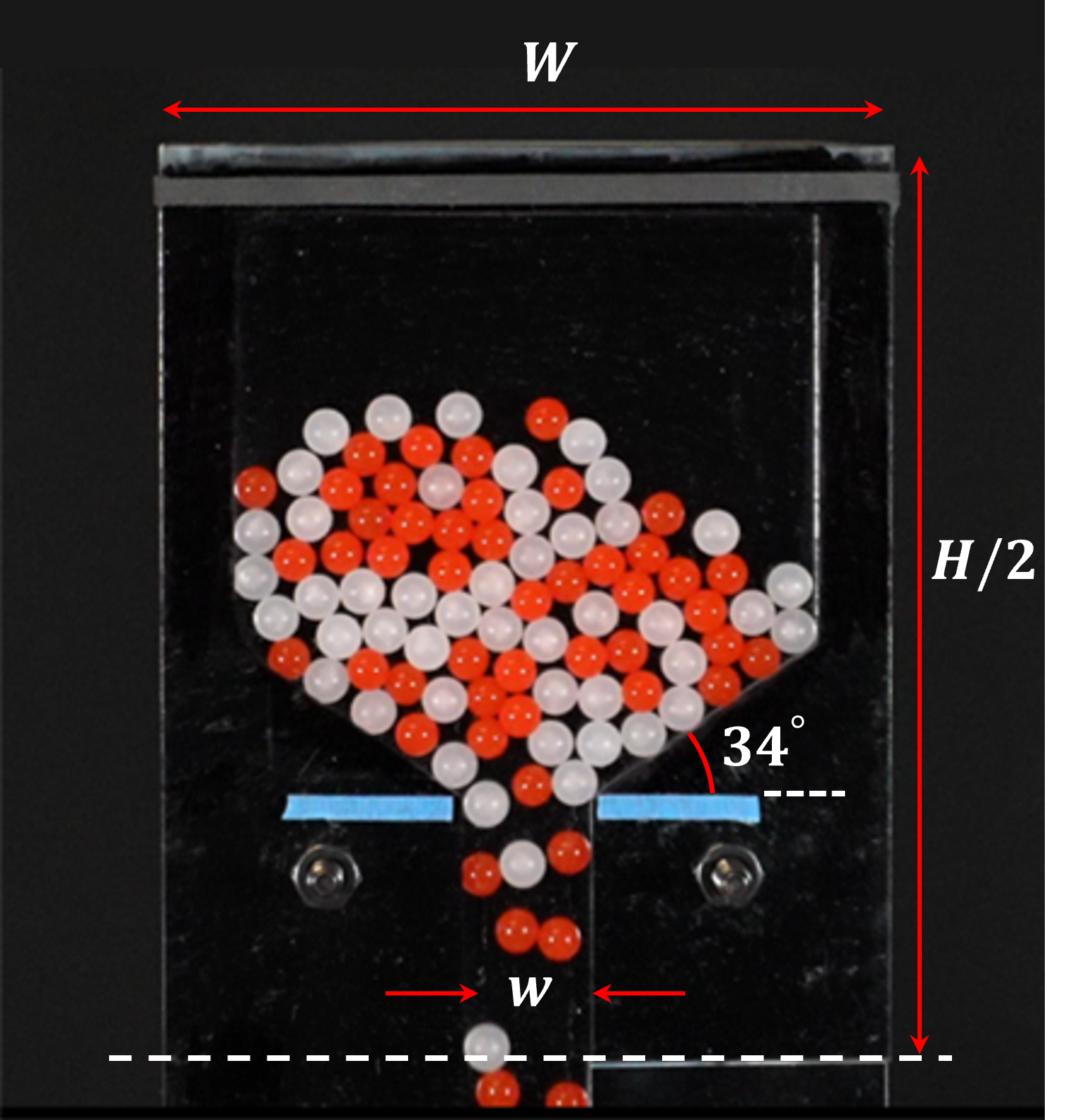}
\caption{Actual image of the working hopper. White particles are rigid and made of polypropylene. Red beads are soft hydrogels. The sizes of all particles are quite similar to avoid any 3D effects. The thickness of the hopper is adjusted to ensure that the particles flow smoothly in a single layer. }
\label{figs:1} 
\end{figure}

\vspace{1em}
We investigate gravity-driven flow of granular mixtures through a custom-built rotating hopper \cite{alborzi2022soft, alborzi2023mixing}. A two-dimensional hopper with dimensions $W\times H = 12.5$cm $\times$ $30.5$cm, a thickness ranging from $9.5$ mm to $12.7$ mm, and a neck width of $w=2.22$ cm is made from acrylic sheets and filled with granular particles, as shown in Fig. \ref{figs:1}. This experiment involves two distinct particle types: soft hydrogel beads composed of polyacrylamide with an elastic modulus 35 kPa and rigid polypropylene spheres. The rigid particles possess a friction coefficient of $\mu_r = 0.21 \pm 0.14$, whereas the soft ones are almost frictionless \cite{alborzi2022soft}. In each hopper flow experiment, the soft and rigid particle sizes are equal. The important geometric parameter is therefore the aspect ratio $w/d$. Three particle sizes are used: $d=7.9$, $9.5$, and $11.1$ mm. To ensure a consistent fill height in the hopper, we carefully select the number of particles across various sizes. Thus, the total number of particles for the three particle sizes of $w/d = 2.80$, $2.33$, and $2.00$ are $126$, $84$, and $60$, respectively. While keeping the total number of particles the same, we vary the number of rigid and soft particles in the mixture, denoted by the rigid fraction $\phi_r$. In this work, we investigate seven different $\phi_r$ values ranging from 0 to 1 for each particle size. 

To calculate the amount of adhesion between different soft-soft, soft-rigid and soft-acrylic contacts, we perform JKR tests \cite{johnson1971surface} using dynamic mechanical analysis (DMA). 
We measure $\sim 13$ mm particles for each contact type and quantify the force and contact radius while the surfaces are pushed against each other. We use a RSA-G2 Solid Analyzer (TA instruments) and attach each object, a spherical hydrogel or acrylic plate, to the top and bottom flat plates of the device. The top tool moves down at a $0.1$ mm/s speed and the force data is collected at 10 Hz. The bottom plate remains fixed. A camera records the experiment at 120 fps to further measure the contact radius during the process.


A single trial consists of the filled hopper undergoing a 180-degree rotation driven by a stepper motor \cite{alborzi2022soft, alborzi2023mixing}. Once the hopper is in the vertical position, it is allowed to rest for a period of time to allow the particles to flow out. The length of each trial is 4 s for the largest particles and 6 s for the other two sizes, as the larger particles drain fastest. Furthermore, we chose the duration of the trials carefully, so that we are confident that any remaining clogs at the end of the trial are permanent clogs.  At the end of the trial the motor applies a small oscillation to the upright position of the hopper to dislodge any particles remaining stuck in an arch. This process is then repeated numerous times to collect sufficient statistical data. The total number of trials at different particle sizes $w/d$ $=$ $2.80$, $2.33$, and $2.00$ are $100$, $100$, and $150$, respectively. We conduct more trials on the condition with the largest particles to ensure sufficient data considering the fewer number of particles.

We record the entire experiment using a Sony $\alpha$7s mirrorless camera capable of capturing footage at 120 fps. Subsequently, we process the recorded videos using custom-written code to extract the desired data. The code analyzes each frame, starting before the first particle exits the hopper. Given that the hopper is still in motion at this point, we dynamically detect the neck area's position.

For this purpose, we affix two blue strips to the front plate, running alongside the exit line and its sides. The code identifies these blue strips and calculates their angle relative to the horizon to determine the hopper's orientation. It then crops the image to a fixed-size rectangle around the neck, ensuring the entire funnel area is visible. The next step involves color masking to differentiate between soft and rigid particles, using red and white colors as reference points.

To locate and track the particles in each frame, we employ an algorithm based on the Hough transform. This information is then fed into a tracking algorithm responsible for determining particle trajectories around the neck. The tracking code follows the principle of minimizing the total squared displacement \cite{crocker1996methods}. It begins by establishing a cutoff radius around each particle in frame $i$ and searches for matching candidates within this radius in frame $i+1$. Subsequently, it considers all possible combinations of $m$ candidates for the $n$ particles visible between the two frames, selecting the specific arrangement that results in the minimum total squared displacement. The algorithm also has a ``memory'' feature that allows particle trajectories to be calculated even if a particle is lost or misidentified in up to 3 frames. 

This process places significant computational demands, necessitating careful selection of the cutoff radius to balance responsiveness and accuracy.  We must ensure that the cutoff radius used by the code remains greater than the actual maximum displacement of a particle as it moves from frame to frame.  In some cases when particles are moving too quickly, the tracking algorithm has difficulty capturing them.  However these errors are limited to high $\phi_r$ situations.

Utilizing the obtained particle trajectories, we track each particle's exit from the hopper by monitoring its passage across the line along the blue strips. The number of remaining particles can be calculated by subtracting the number of exited particles from the total.  Tracking also enables us to calculate particle velocities and time intervals between consecutive particle exits.  


\vspace{1em}
\section{Results}
\vspace{0.5em}

\vspace{2em}
\subsection{V\lowercase{elocity}}
\vspace{1em}

\begin{figure}[!h]
    \centering
    \begin{minipage}{.49\textwidth}
        \centering
        \includegraphics[width=\linewidth]{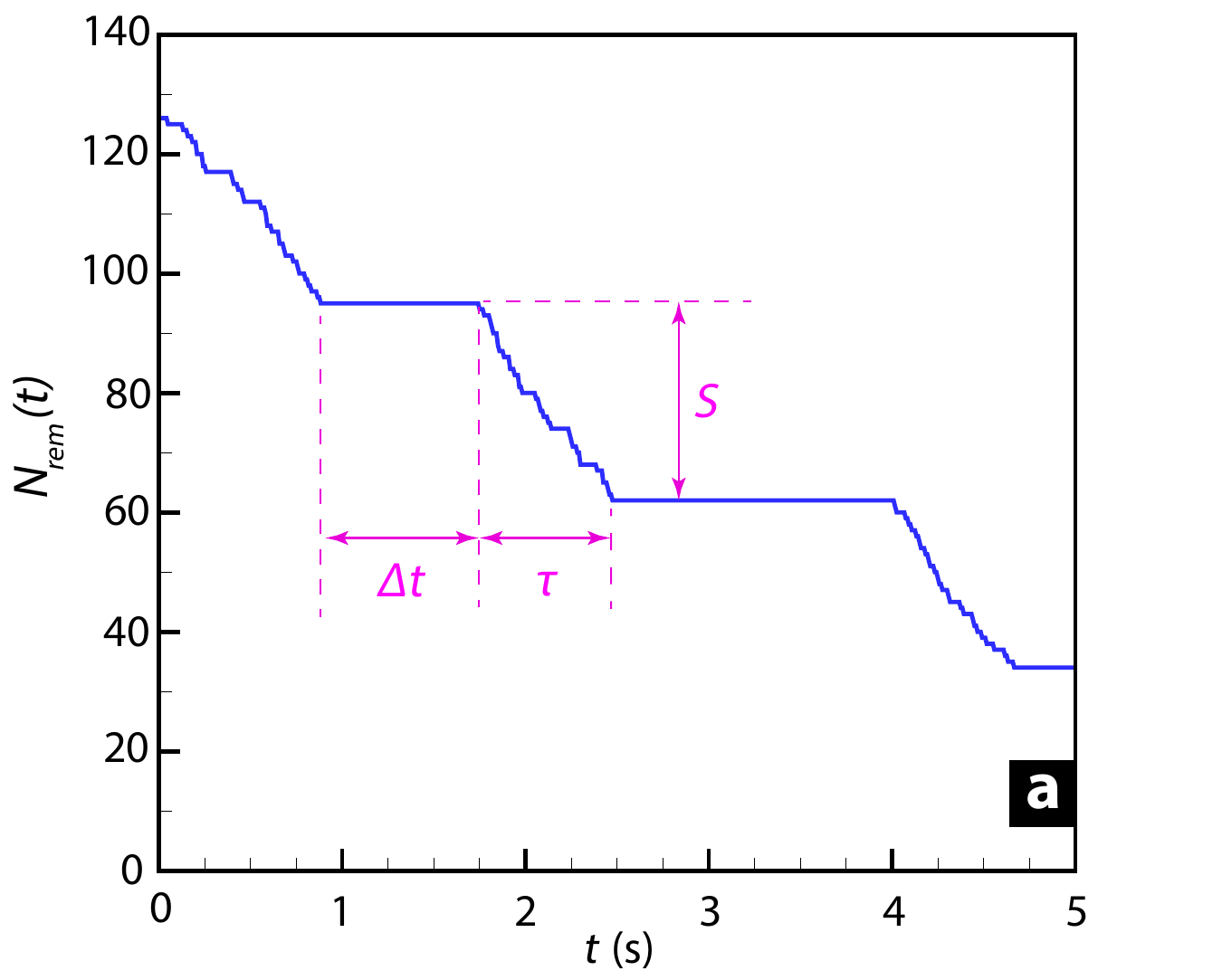}
    \end{minipage}%
    \begin{minipage}{.49\textwidth}
        \centering
        \includegraphics[width=\linewidth]{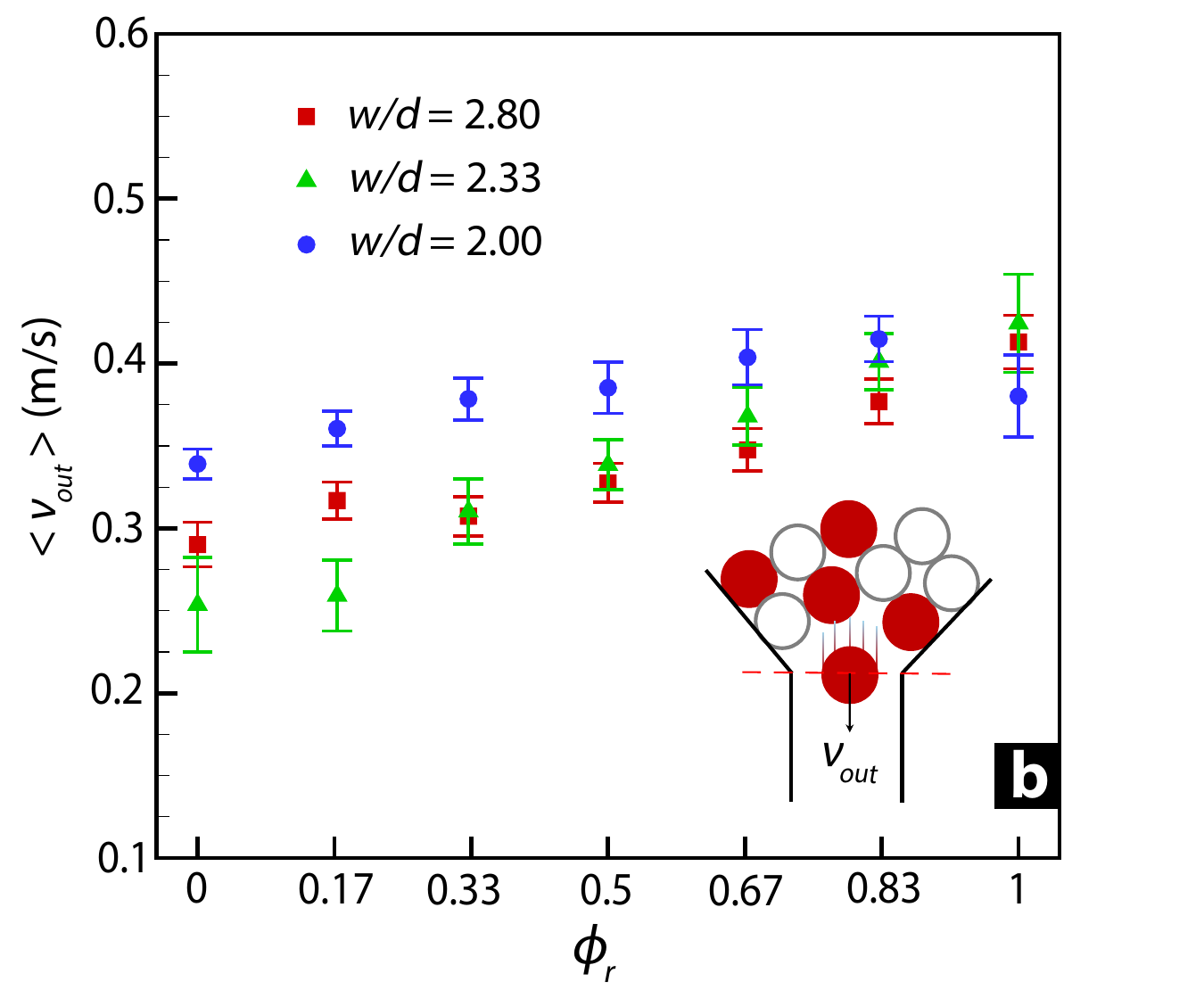}
    \end{minipage}
    \begin{minipage}{.49\textwidth}
        \centering
        \includegraphics[width=\linewidth]{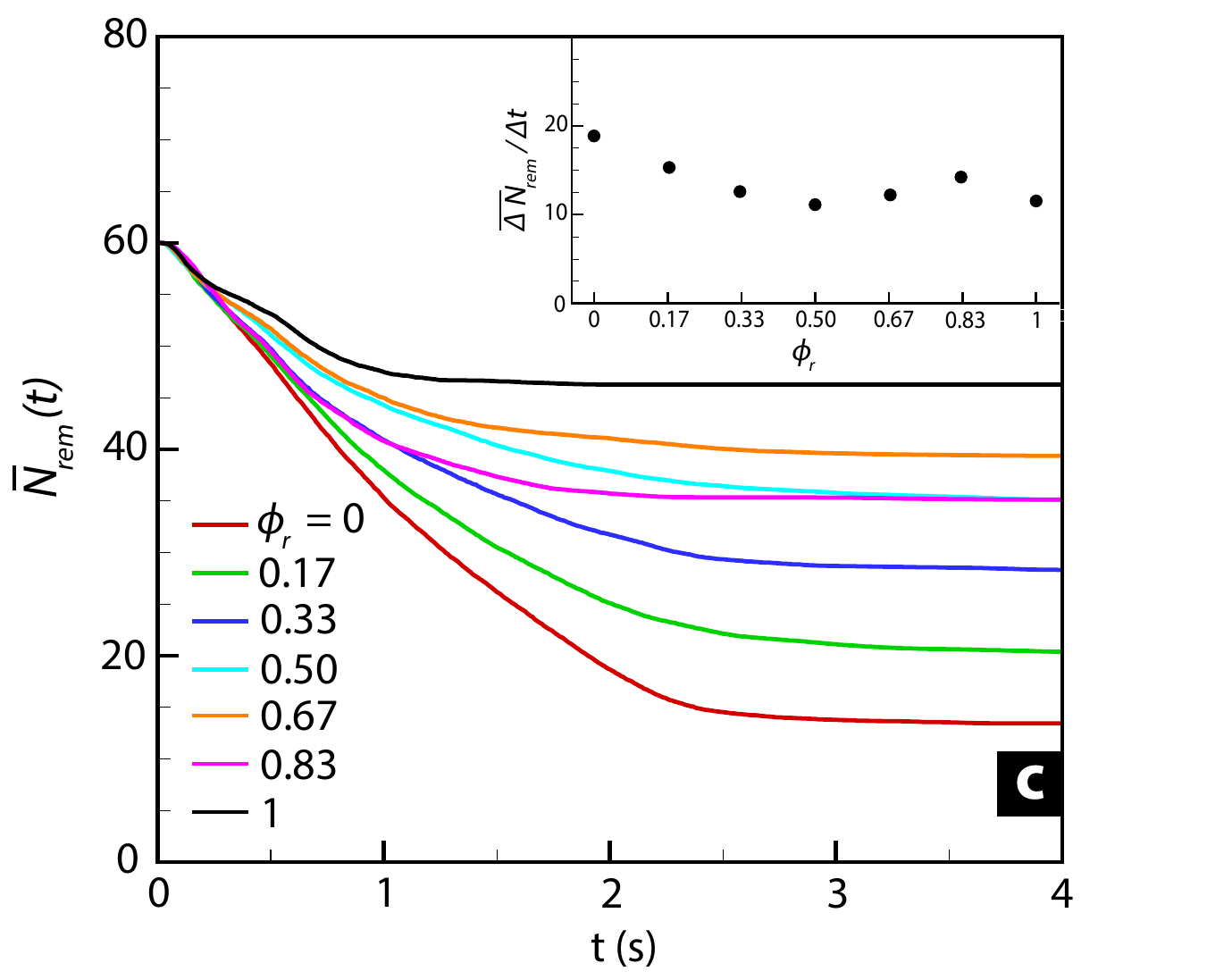}
    \end{minipage}
  \caption{(a) The number of remaining particles in the hopper over time, during a single trial at $w/d=2.33$ and $\phi_r=0.67$. $S$ is the avalanche size, $\tau$ the avalanche duration, and $\Delta t$ the arrest time. (b) The average steady velocity of individual particles at the exit as a function of rigid fraction for three different particle sizes. Each point represents the average of all exiting particles in all trials at a given $\phi_r$. Error bars are the standard deviation. (c) The number of remaining particles as a function of time for $w/d=2.00$, averaged over 150 trials. The inset shows the average slope of the linear portion before the plateaus.}
  \label{figs:2}
\end{figure}

In every hopper rotation, the particles start to fall down the neck until completely drained or clogged. The final state of the hopper depends on whether a stable arch is formed at the neck or not. Earlier studies show clogging probability increases with rigidity, surface friction, rigid fraction of the sample, and particle size \cite{alborzi2022soft}. During the flow, however, the particles can form arrangements that temporarily stop flowing and restart spontaneously, a phenomenon known as an \textit{avalanche}. The presence and duration of temporary clogs can be quantified by examining the outflow rate of particles exiting the hopper.  Fig. \ref{figs:2}a shows the number of particles remaining in the hopper over time, $N_{rem}(t)$, for a system in which $w/d=2.80$ and $\phi_r=0.67$. Within the first $\sim 0.9$ s, $N_{rem}(t)$ starts to decrease from the initial full state of $N_0 = 126$ particles. The plateau beginning at $t\sim 0.9$ s indicates a temporary clog that endures for nearly $\Delta t=0.85$ s. An avalanche then occurs that lasts for $\tau = 0.75$ s and ends with another temporary clog. During the avalanche, the outflow rate, or the slope of $N_{rem}(t)$, is similar to that of early times. A number of $S=33$ particles drain between the two temporary clogs which is referred to as \textit{avalanche size}. Additional shorter plateaus are visible throughout the entire discharge. Approximately 34 particles remain in the hopper at the end of the trial. 

To evaluate how fast individual rigid or soft particles move out of the hopper, we examine the vertical component of their outflow velocity at the exit, shown in Fig. \ref{figs:2}b. The velocity of each exiting particle, $v_{out}$, is measured as it passes a horizontal line just below the hopper exit, as shown in the inset schematic.  Fig. \ref{figs:2}b shows the average outflow velocity of individual particles, $<v_{out}>$, calculated by taking an average over all particles at the exit in all trials of a certain condition. That is, each data point represents an average velocity for $N\sim O(10^4)$ particles. All particles, upon exiting the hopper, fall at velocities between 500-800 mm/s.  As shown in Fig. \ref{figs:2}b, when rigid fraction increases, average particle velocity increases too. Rigid particles move faster because they do not undergo any deformation when impacting the walls, hence preserving more of their kinetic energy after the impact. The same phenomenon applies to the particle-particle interactions. Rigid particles tend to bounce back faster after colliding with another rigid particle compared to a soft-soft collision. In addition, larger particles appear to fall faster because there are generally fewer of them inside the hopper.  As such, larger particles have more room to fall from relatively higher initial positions. Here we note that particle velocity during discharge is not quite steady over the entire duration of each trial.  At early times $v_{out}$ is faster due to the hopper rotation.  $v_{out}$ slows below its steady-state value towards the end of a trial due to the reduced fill height of the hopper. An example of this behavior is shown in Fig. S1, where $\overline{v}_{out}$ at certain values of $\phi_r$ is plotted against $N_{exited}$ for $w/d=2.80$. $\overline{v}_{out}$ represents the average value of individual outflow velocities between all trials at a certain $\phi_r$ and $N_{exited}$. Even so, nearly 80\% of the particles in each run leave the hopper at the steady velocity.



The individual particle velocities, however, do not predict the overall discharge rate of the hopper.  In referring to the discharge rate, we are interested to see how quickly or slowly a full hopper becomes empty. There might be temporary or permanent clogs during a particular trial which increase the overall discharge time. Thus, in addition to measuring individual particle velocities, we also investigate average \deleted{discharge}\added{outflow} rates. To obtain the average \deleted{discharge}\added{outflow} rate \deleted{$\bar {N}_{rem}$}\added{$|d \bar{N}_{rem} /dt|$}, we \added{first} take the average, over all trials, of the instantaneous number of particles remaining in the hopper at every moment during the experiments. The results are shown in Fig. \ref{figs:2}c for all $\phi_r$ values at $w/d=2.00$. For all $\phi_r$, $\bar{N_{rem}}$ decreases smoothly and then levels out to a plateau. At very short times, $t\lesssim0.2$ s, $\bar{N}_{rem}(t)$ collapses to a single curve for all values of $\phi_r$, before any significant clogging occurs. The long-time plateau in $\bar{N}_{rem}$ indicates the average number of particles remaining in the hopper at the end of each trial. At the end of each trial, $\bar{N}_{rem}$ increases as $\phi_r$ increases, from 13 when $\phi_r=0$ to 46 when $\phi_r=1$. This increases matches the observation that clogging probability also increases with $\phi_r$ \cite{alborzi2022soft}. The increasing number of clogs also determines the time at which the plateau is reached.  That is, when $\phi_r=1$, $\bar{N}_{rem}$ reaches a plateau in approximately 1 s (in black). In contrast, when $\phi_r\leq 0.5$, more than $\sim2.5$ s elapse before $\bar{N}_{rem}$ reaches its plateau. Before the final plateau is reached, the slope of $\bar{N}_{rem} (t)$ also depends on $\phi_r$. That is, the outflow rate $|d \bar{N}_{rem} /dt|$ generally increases as $\phi_r$ decreases, as shown in Fig. \ref{figs:2}c inset. For instance, $|d \bar{N}_{rem} /dt|\sim10$ particles/s when $\phi_r=0$, while $|d \bar{N}_{rem} /dt|\sim25$ particles/s when $\phi_r=1$.


The average amount of particles that exit before a permanent clog occurs, $\overline{\Delta N}_{rem}=N_{rem}(t=0)-N_{rem}(t\rightarrow\infty)$, is plotted in Fig. S2 against $\phi_r$ for the data shown in Fig. \ref{figs:2}c. Interestingly, \deleted{faster outflow rates}\added{larger $\overline{\Delta N}_{rem}$} are observed when soft particles dominate the mixture, even though soft particles, individually, move more slowly than rigid ones. That is, \deleted{discharge rates are}\added{the average number of discharged particles is} not \deleted{equivalent}\added{positively correlated} to particle velocities because they reflect the probability that clogs will form.  Even short-lived, temporary clogs cause $|d \bar{N}_{rem} /dt|$ to slow. \deleted{Rigid particles fall faster, which gives rise to higher clogging probabilities: particles have less time to escape out of a potential clog before hitting a neighbor. Therefore}\added{Due to particle rigidity and higher friction}, more permanent clogs  occur when rigid particles dominate the mixture.  Thus, the overall \deleted{discharge rate}\added{number of discharged particles} decreases. \deleted{This observation supports the notion of ``\textit{faster-is-slower}'' seen elsewhere in the literature \cite{pastor2015experimental}.}




\vspace{1.5em}

\subsection{T\lowercase{ime intervals}, $\Delta t$}
\vspace{1em}

\begin{figure}[h]
\centering

\includegraphics[width=0.49\columnwidth]{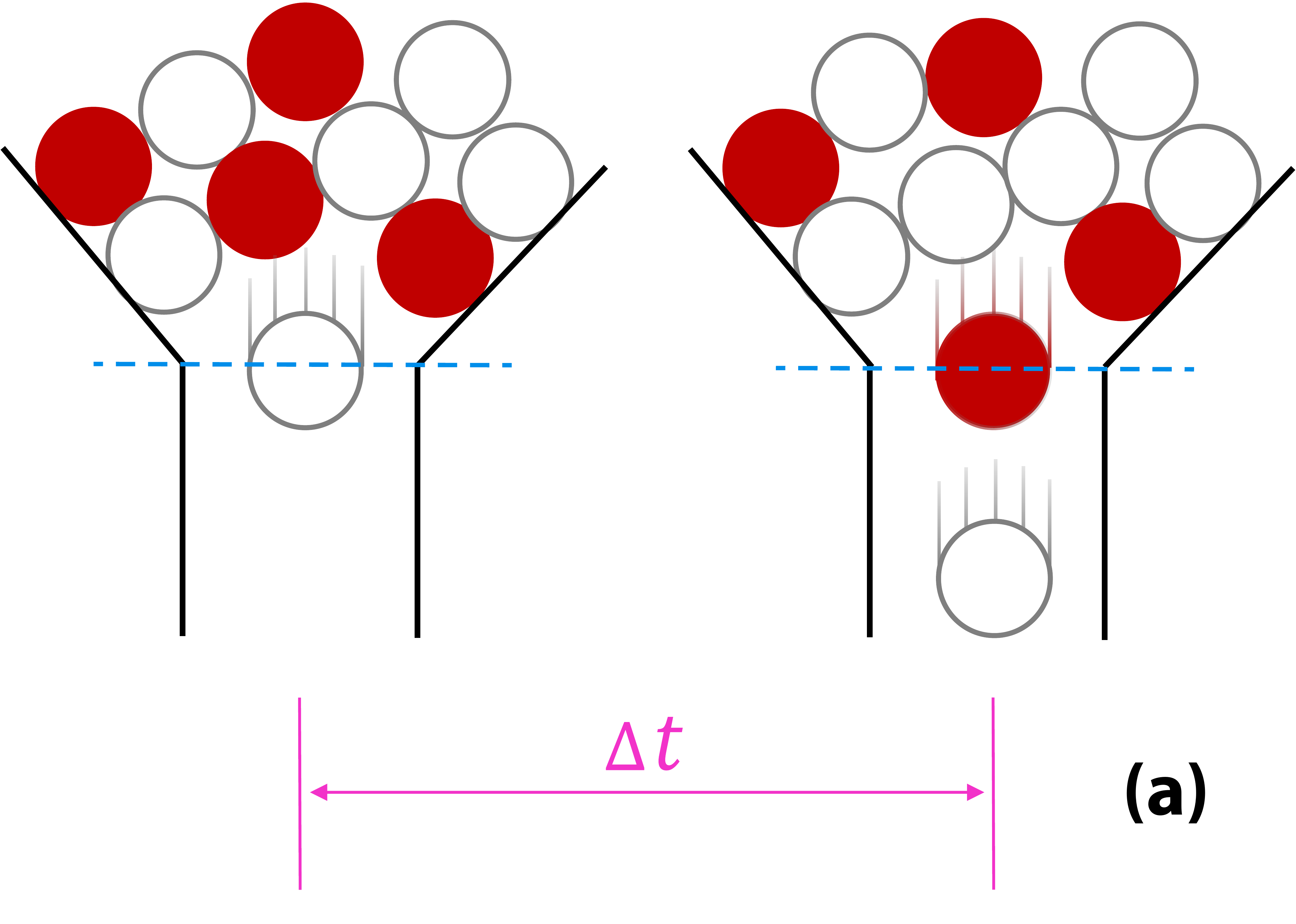}
\includegraphics[width=0.49\columnwidth]{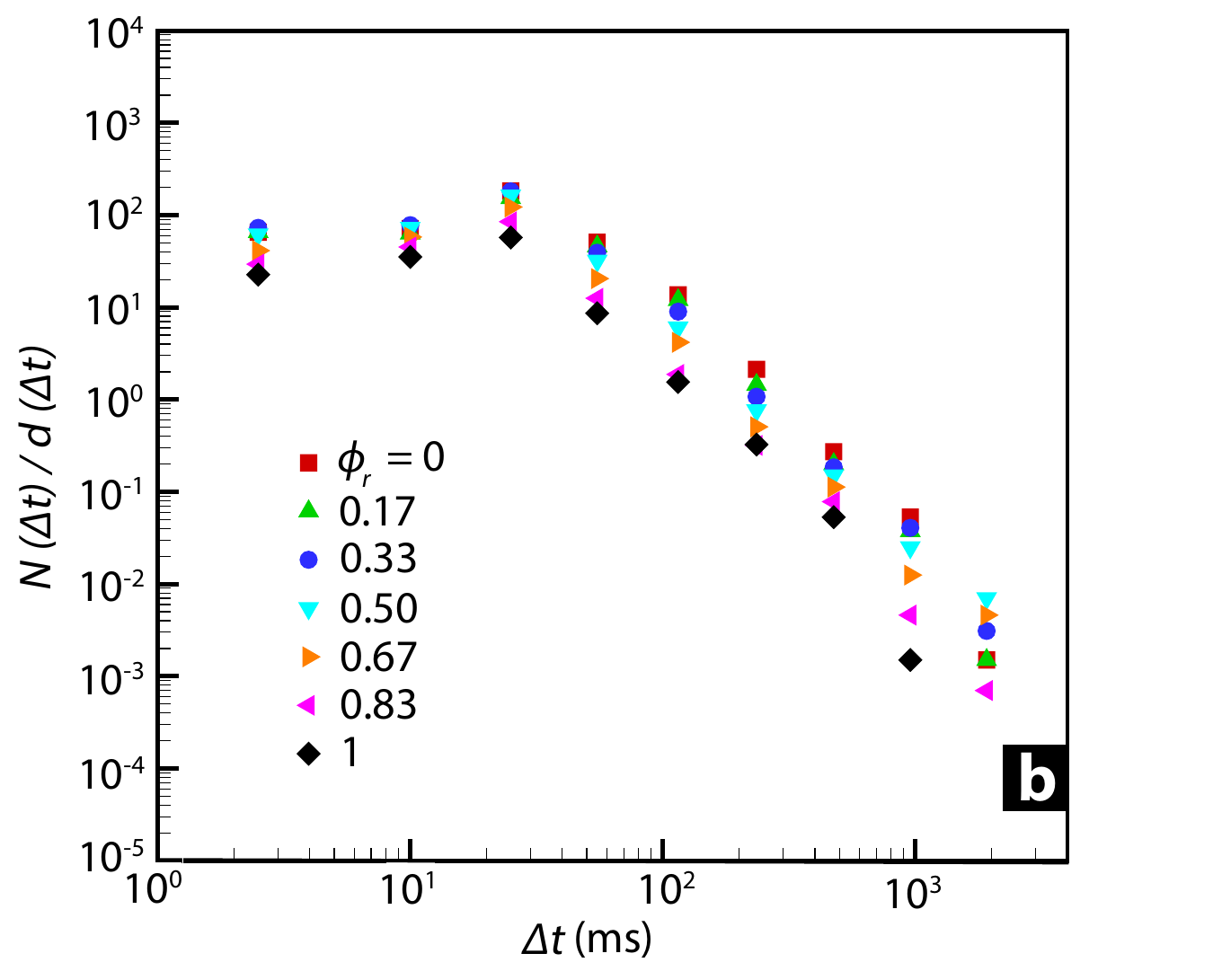}
\caption{(a) Illustration of $\Delta t$, the time lapse between the consecutive exiting particles. (b) Distribution of $\Delta{t}$ for $w/d=2.33$, exhibiting a power-law tail $(\Delta t)^{-\alpha}$ with exponent $\alpha = 2.56$.}
\label{figs:3}
\end{figure}

The probability distribution function of time intervals $P(\Delta t)$ has been also a point of interest in similar studies. In the current study, in contrast to others, we investigate mixtures of soft and rigid particles in addition to systems that are uniformly soft or rigid. We measure $\Delta t$ between all consecutive particle exits, as shown in Fig. \ref{figs:3}a, occurring in all 100 trials of each system composition.  A particle exit is defined at the time at which the particle falls below a fixed horizontal position.  $\Delta t$ is also indicated in Fig. \ref{figs:2}a. Fig. \ref{figs:3}b shows the log-log plot of the distribution $N(\Delta t)$ at all rigid fractions for the case $w/d=2.33$, where the $x$-axis is logarithmically binned in order to provide better visualization of the tail of the data \cite{newman2005power}. The $y$-axis values are then normalized by the corresponding bin width at each point. Given the number of particles in the hopper and the total number of trials, the distributions shown in Fig. \ref{figs:3}b represent consecutive particle exits numbering between $N\sim2000$ at $\phi_r=1$ and $N\sim8000$ at $\phi_r=1$.  $N$ decreases with $\phi_r$ due to the increased probability of clogs at the end of each trial \cite{alborzi2022soft}.  For $\Delta t< 25$ ms, numbers grow with a power-law of slope $0.40$ until they peak at $\Delta t = 25$ ms. This short-time regime corresponds to freely falling particles with $\Delta t$ less than four frames of the video. Beyond $\Delta t = 25$ ms, the frequency of waiting time between consecutive particle exits decreases with a second power-law with exponent $\alpha=2.56$ for $\Delta t \geq 25$ ms.  Interestingly, this dual behavior persists regardless of $\phi_r$.  For each system of particles, the most likely waiting time is $\Delta t = 25$ ms. Longer waiting times between particle exits become increasingly less frequent. The existence of the short-time growth regime for $\Delta t$ is not widely reported, to the authors' knowledge. In microfluidic hopper flows of emulsions, $P(\Delta t)$ follows an exponential distribution when the flow is smooth and non-intermittent \cite{hong2022clogging}. Intermittent flow in the same emulsions exhibits a decaying power-law for $P(\Delta t)$, as observed in the current study. However, both the emulsion flow rate and image capture rate are very slow, and short time delays between particle exits may not be fully resolved \cite{hong2022clogging}. 


Here, it is common to define a threshold $\Delta t$ to differentiate the breakage of temporary clogs from free flowing particles. Video1 in the supplemental materials provides examples of consecutive falling particles where $\Delta t = 83$, 125, 167 and 208 ms, corresponding to $\Delta t = 10$, 15, 20, and 25 frames.  Our observations indicate that the flow is noticeably paused when $\Delta t > 125$ ms. By eye, $\Delta t$ values less than 125 ms do not appear to be flow stoppages.  Therefore, we define 125 ms as the \textit{avalanche threshold}, $\Delta t_c=125$ ms. That is, any plateau in $N_{rem}(t)$, as in the example in Fig. \ref{figs:2}a, that lasts longer than $\Delta t_c=125$ s is considered a temporary clog or arch that breaks to form an avalanche. Interestingly, this threshhold 125 ms falls within the power law decay region of Fig. \ref{figs:3}b.

While Fig. \ref{figs:3}b shows the frequency of waiting times between particle exits, it does not provide any information about when, during the course of a particular flow test, these exits occur.  Therefore, we also plot $\Delta t$ for each exiting particles over the entire duration of the all trials.  Fig. S3 shows scatter plots of $\Delta t$ as a function of $N_{exited}$ for three different values of $\phi_r$ at $w/d=2.33$. Longer waiting times are seen early in each trial when $\phi_r=1$: $\Delta t > \Delta t_c$ occurs mainly for the first $\sim20$\% of exiting particles. This may indicate traffic jams as fast moving rigid particles try to exit. In contrast, longer waiting times are seen toward the end of each trial when $\phi_r=0$, which may indicate the slowness of few remaining soft particles that exit after sliding down the side wall. Interestingly, the intermediate value of $\phi_r=0.5$ exhibits the longest waiting times $\Delta t$.  These mainly occur in the middle of the trials, as soft and rigid particles compete to exit the hopper.  A similar situation is seen in systems with $w/d=2.00$ and 2.80.


In gravity driven flows, the occurrence of avalanches has been found to depend on particle softness \cite{harth2020intermittent}. Spontaneous avalanches occur only if the particles are sufficiently soft to enable intermittent outflow conditions. To understand the avalanche frequencies as a function of the rigid fractions in our mixed systems, we quantify the number of temporary clogs at every $\phi_r$, where a temporary clog has a waiting time $\Delta t >\Delta t_c$. However, any clogs or arches present at the end of a trial are considered to be permanent clogs, and should be excluded from this analysis. Thus the upper limit on $\Delta t$ is written as $\infty$. The probability of temporary clogs or, equivalently, the probability of avalanches can be calculated from the following:


\begin{equation}
    P (\Delta t_c < \Delta t < \infty  ) = \frac{N (\Delta t_c < \Delta t < \infty )}{N_t(\Delta t)}
    \label{eq:1}
\end{equation}

\noindent The denominator $N_t(\Delta t)$ represents all time intervals $\Delta t$ between consecutive falling particles.  The numerator represents the number of those time intervals that is greater than 125 ms, while excluding clogs present at the end of a trial.  As such, the avalanche probability $P(\Delta t_c < \Delta t < \infty)$ represents the fraction of consecutive falling particles which have waited longer than 125 ms after the previous particle fell.

\begin{figure}[H]
\centering
  \includegraphics[width=0.6\columnwidth]{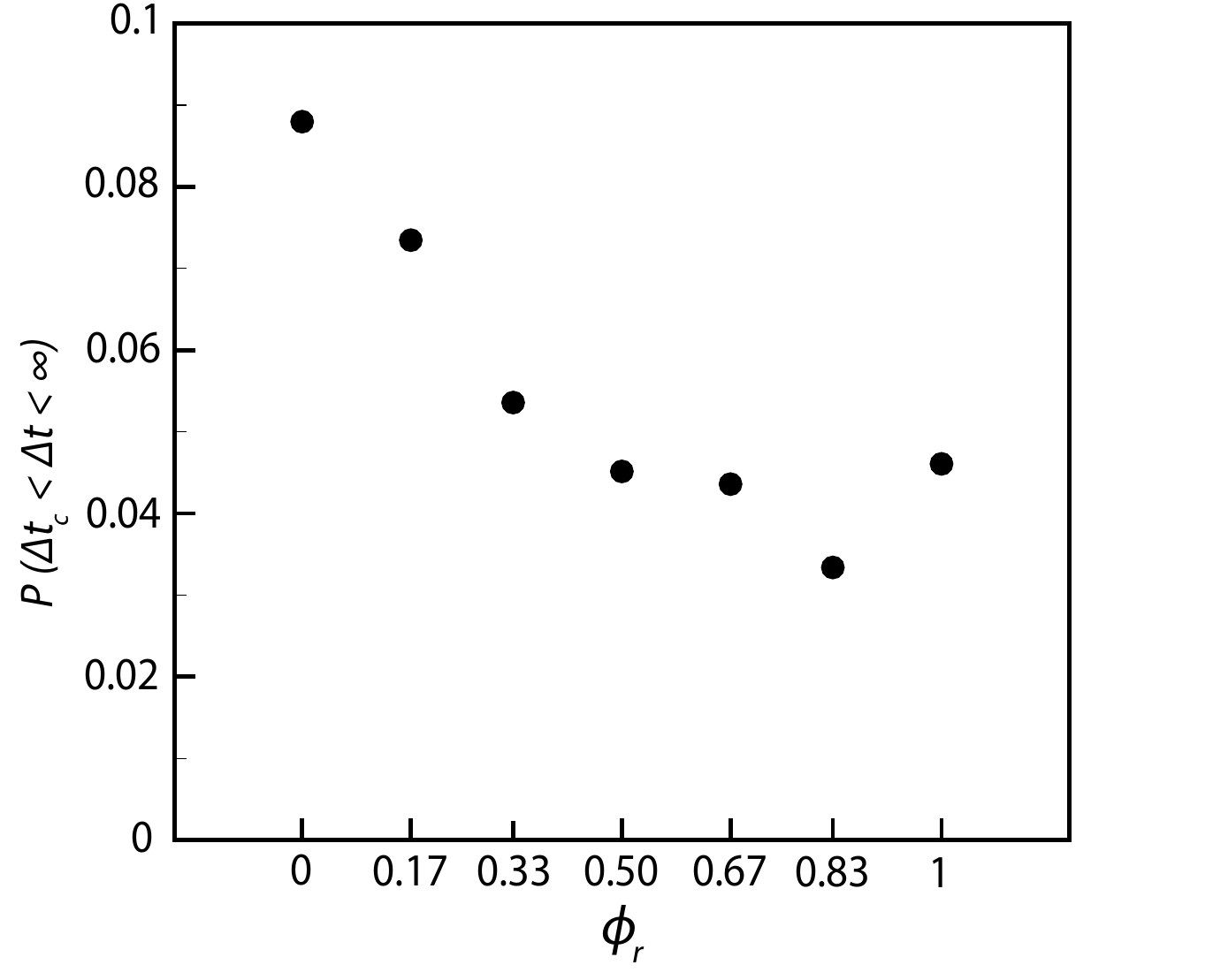}
  \caption{Probability of transient flow interruptions calculated from Eq. \ref{eq:1} with the data in Table \ref{table:1} at $w/d=2.33$. The critical time interval, $\Delta t_c=125$ ms, defines the threshold below which only free particle flow occurs.}
  \label{figs:4}

\end{figure}

Table \ref{table:1} indicates the statistics used to calculate $P(\Delta t_c < \Delta t < \infty)$ for every particle size.  That is, for each size ratio $w/d$, $N_t(\Delta t)$ indicates the total number of time lapses between the consecutive particle exits, and $N(\Delta t_c<\Delta t<\infty)$ the number of time lapses longer than $\Delta t_c$, the threshold that defines a pause in the flow.  The results of $P(\Delta t_c < \Delta t < \infty)$ are shown in Fig. \ref{figs:4} for the specific particle size $w/d=2.33$, where the number of events are sufficient to see the behavior variations. The avalanche probability decreases as $\phi_r$ increases.  The apparent non-monotonic behavior at $\phi_r=1$ can be explained by the analysis protocol. The tracking algorithm can fail to track the fastest-falling rigid particles and therefore $N_t(\Delta t)$ is slightly underestimated, which results in a minor growth of $P$ at $\phi_r=1$. The overall decrease in $P$, however, shows that avalanches happen less frequently as the system consists of more rigid particles. Rigid particles flow faster than soft particles and are only stopped by permanent clogs. In contrast, soft particles flow more slowly, but intermittently get trapped by temporary clogs which disrupt spontaneously and the particles resume the flow.


\begin{table}[h]
\centering
\caption{Number of total time lapses (particle exits) $N_t(\Delta t)$ and the time lapses greater than the critical value $\Delta t_c.$ (number of arrest events).}
\label{table:1}
\begin{tabular}{c c c c c c c}
\hline \\ [-5pt]
& \multicolumn{2}{c}{$w/d=2.80$} & \multicolumn{2}{c}{$w/d=2.33$} & \multicolumn{2}{c}{$w/d=2.00$} \\[-5pt] &\multicolumn{2}{c}{\rule[0pt]{100pt}{0.5pt}} &\multicolumn{2}{c}{\rule[0pt]{100pt}{0.5pt}}&\multicolumn{2}{c}{\rule[0pt]{100pt}{0.5pt}}\\ [2pt]
 {$\phi_r$} & {\shortstack[c]{$N_t(\Delta t)$}} & {$N(\Delta t_c<\Delta t<\infty)$} & {\shortstack[c]{$N_t(\Delta t)$}} & {$N(\Delta t_c<\Delta t<\infty)$} & {\shortstack[c]{$N_t(\Delta t)$}} & {$N(\Delta t_c<\Delta t<\infty)$} \\ [5pt] \hline \\ [-5pt]

   0 & 11842 & 155 & 8275 & 728 & 6835 & 169 
   \\ 0.17 & 11459 &146 & 7146 & 525 & 5794 &	210
 \\ 0.33 &10849 & 182 & 7317 &392 & 4605 &188
 \\  0.50 & 9887 &199 & 6247 &282 & 3593 &177
\\  0.67 & 7103 &173 & 4520 &197 & 2948 & 161
 \\  0.83 & 5258 &111 & 3026 &101 & 3581 &139
 \\  1 & 4211 &57 & 2149 &99 & 2595 &354 \\

\end{tabular}

\end{table}

Fig. \ref{figs:3} indicates that the presence of soft particles in the mixture causes a larger number of consecutive exits for all values of $\Delta t$, the interval between particle exits.  That is, the normalized measurements of $N(\Delta t)$ for $\phi_r \leq 0.5$ all lie above the measurements of $N(\Delta t)$ for $\phi_r > 0.5$.  This trend holds for all $\Delta t$ except the very longest, when $\Delta t > 1$s.  However, at long $\Delta t$ values, the number of total consecutive exits decreases due to the increase in permanent clogs.  Fig. \ref{figs:4} indicates that the probability of pauses long enough to signify flow stoppage decreases with the rigid particle fraction.  Soft particles cause more flow pauses.



Underlying the results shown in Fig. \ref{figs:3}, Fig. \ref{figs:4}, and Table \ref{table:1} is an increase in the number of permanent clogs that occurs as $\phi_r$ increases.  As $\phi_r$ decreases, however, the number of intermittent clogs increases.  The increase in permanent clogs can be explained, in part, by frictional contacts between rigid particles \cite{alborzi2022soft}.  However, friction is absent in contacts between soft particles.  Therefore, we do not appeal to friction to describe the increase in temporary flow pauses that occurs when soft particles predominate the mixtures.  Instead, we investigate the adhesion between particles as a possible explanation of temporary pauses in flow.

\begin{figure}[h]

\centering
  \includegraphics[width=\columnwidth]{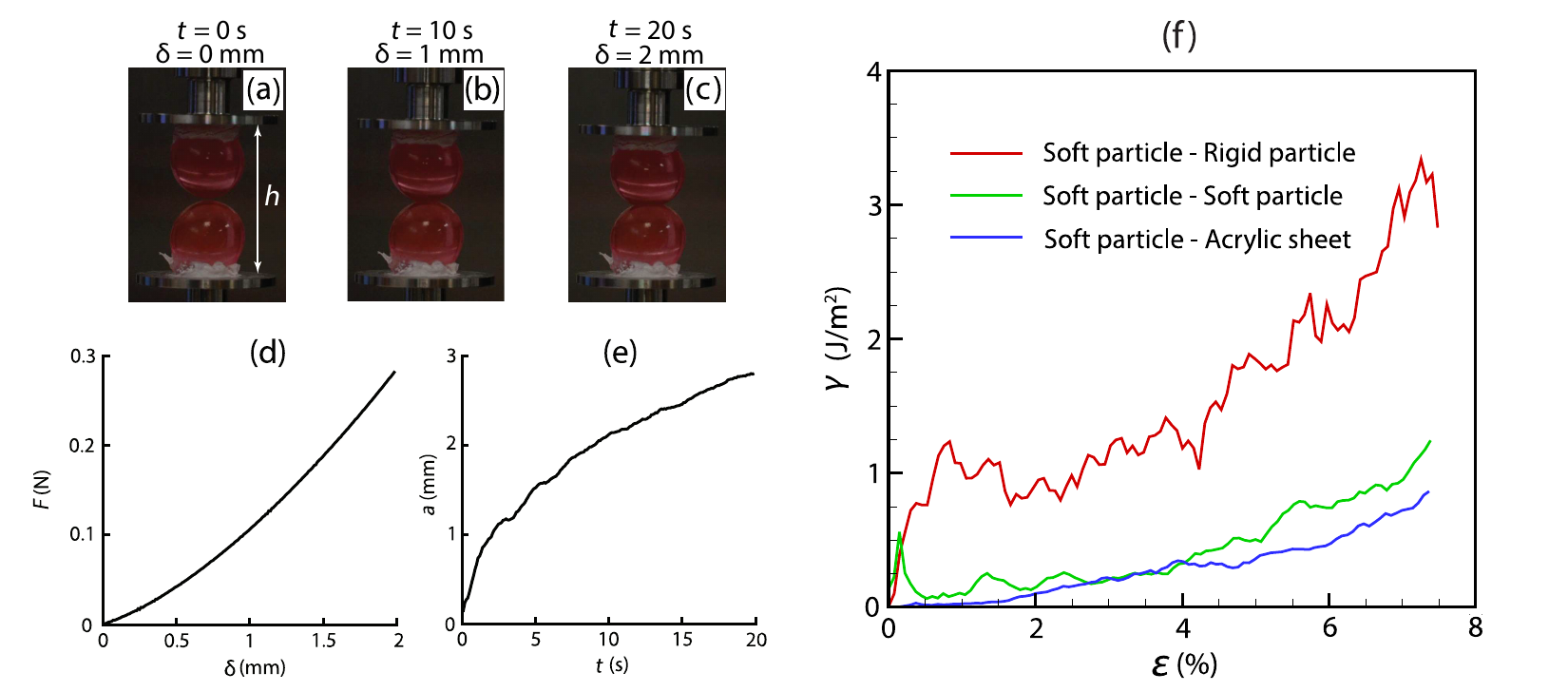}
  \caption{(a)-(c) Snapshots from the first, middle, and last stages of the soft-soft JKR test with particle size $d=13.4$ mm. (d) Force vs. displacement and (e) contact radius variation with time corresponding to the same soft-soft test. (f) Adhesion energy against strain, obtained from Eq. \ref{eq:2} for various particle contacts.}
  \label{figs:5}

\end{figure}

Adhesion causes particles to stick together and thus can hinder particle motion. We perform JKR tests to quantify adhesion between each particle type and between the soft particles and the acrylic plates \cite{johnson1971surface}.  Fig. \ref{figs:5} illustrates the steps in the measurement and corresponding analysis.  Figs. \ref{figs:5}a-\ref{figs:5}c show example images collected in a measurement of two soft particles, each glued to the upper and lower plates of the DMA.  The three images correspond to different values of $h$, the gap between the plates, as the upper plate moves downward with increasing displacement $\delta$.  The plots in Figs. \ref{figs:5}d and \ref{figs:5}e show the raw data as collected: the DMA instrument meausures the force $F$ between the particles as a function of $\delta$.  Because the upper plate moves with constant velocity, we can convert $\delta$ to $t$.  Image analysis of the video measures the contact radius $a$ as a function of time.  With both $F$ and $a$ measured as a function of time, we correlate the two independent data sets using the moment of contact.

When the contact radius is small compared to the gap between the plates, $a/h\rightarrow0$, the JKR expression can be used to calculate the adhesion energy:

\begin{equation}
    \gamma = \frac{(4E^*a^3/3R-F)^2}{8\pi E^*a^3}
    \label{eq:2}
\end{equation}

\noindent where $\gamma$ is the adhesion energy in J/m\textsuperscript{2} and $1/R=1/R_1+1/R_2$.  The subscripts 1 and 2 indicate the two materials in contact. The elastic modulus is embedded in the parameter $E^*$ as follows: $1/E^*=(1-\nu_1^2)/E_1+(1-\nu_2^2)/E_2$, where $\nu$ is the Poisson ratio.  For the soft hydrogels, $E_s=35$ kPa and $\nu_s=0.3$ \cite{alborzi2023mixing}. Both the rigid spheres and the acrylic plate have $E\sim O(1)$ GPa, meaning they are 5 orders of magnitude more rigid than the soft particles.  Therefore, the contribution of the rigid material to $E^*$ vanishes.




The calculated adhesion energies for each contact type are shown in Fig. \ref{figs:5}f as a function of strain, $\epsilon$, calculated from the change in particle diameter through the course of the measurement: $\epsilon = \Delta d/d$. As shown, $\gamma$ is largest for the contact between soft and rigid particles throughout the entire range of $\epsilon$.  Contact between two soft particles and between a soft particle and an acrylic surface have nearly equal adhesion energies; both are between $\sim 10-30\%$ of $\gamma$ measured for contact between soft and rigid particles.  Measurements of adhesion between glass spheres and flat elastomeric surfaces suggest a similar order of magnitude in adhesion energy \cite{berman2019singular}.  When a glass bead is pushed into an elastomeric surface with modulus 5 kPa, the measured elastocapillary length is 4 $\mu$m at zero strain, increasing to 10 $\mu$m as strain increases \cite{berman2019singular}.  Because the elastocapillary length scales like $\gamma/E$, this suggests $\gamma \sim 0.02-0.05$ J/m$^2$.  Extrapolating to a substrate with the modulus of the soft particles in this study, $E_s=35$ kPa, increases the adhesion energy to $\gamma \sim 0.14 - 0.35$ J/m$^2$.  Indeed, Fig. \ref{figs:5}f shows that $\gamma$ reaches $\sim 0.3$ J/m$^2$ at $\epsilon \sim 4\%$ strain.

The observation that $\gamma$ for soft-soft contacts is lower than $\gamma$ for soft-rigid contacts may be surprising and counter-intuitive.  However, it may also help explain the occurrence of temporary flow stoppages when soft particles predominate the mixture.  When soft-soft contacts predominate,  only a small amount of adhesion must be overcome to break a temporary arch.  As the number of soft-rigid contacts increases, more adhesion must be overcome.  As rigid-rigid contacts become important, above $\phi_r=0.5$, friction between the rigid spheres also contributes \cite{alborzi2022soft}.  The combination of increasing adhesion and increasing friction reduces the number of temporary flow pauses as $\phi_r$ increases, while the number of permanent clogs increases.

\begin{figure}[h]

\centering
    \includegraphics[width=0.49\columnwidth]{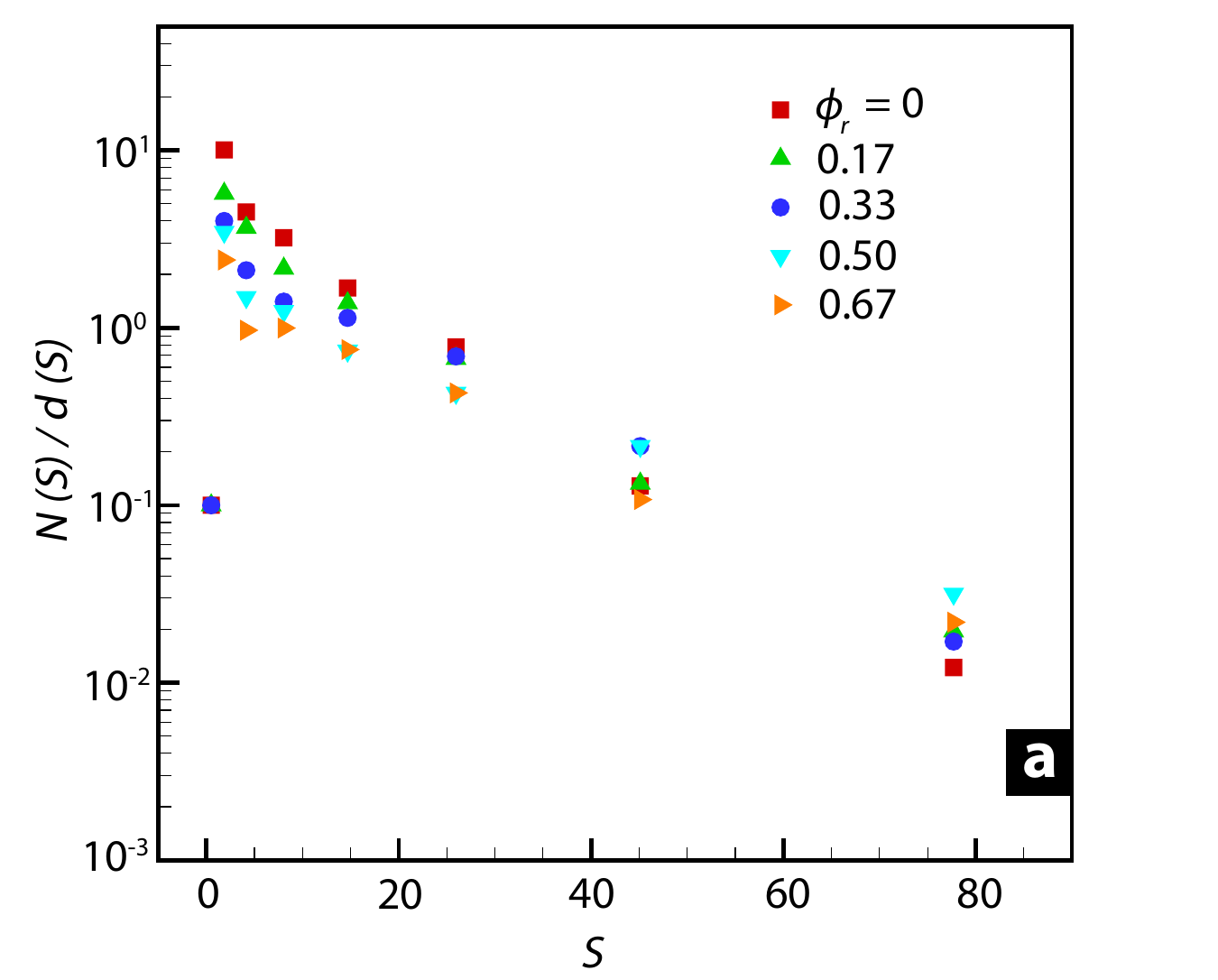}
  \includegraphics[width=0.49\columnwidth]{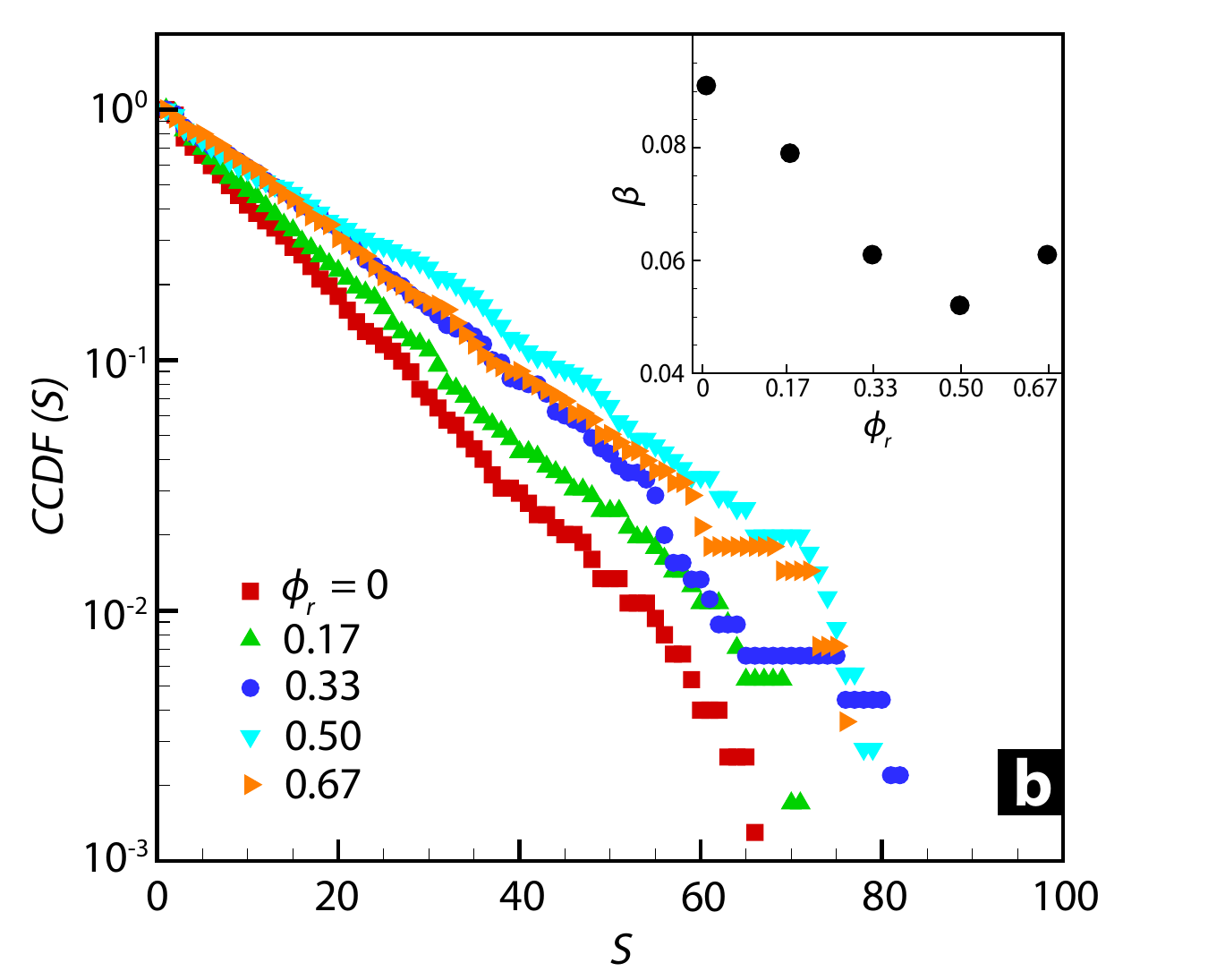}
  \caption{(a) Distribution of avalanche sizes $S$ for $w/d=2.33$ showing an exponential behavior $exp(-\alpha S)$ with average exponent $\alpha=0.07$. The x-axis is logarithmically binned from $S_{n+1}=S_n+1.7^n$ where $0\leq n \leq 8$ and $S_0=0$. (b) The complementary cumulative distribution function of the same data defined by Eq. \ref{eq:3}. The inset shows the corresponding exponents $\beta$ of the exponential fits $exp(-\beta S)$ for each trace.}
  \label{figs:6}
\end{figure}

\vspace{1.5em}

\subsection{A\lowercase{valanche size}, S}
\vspace{1em}


Next, we look at the distribution of avalanche sizes across different rigid fractions. We define the avalanche size $S$ as the number of particles falling between each two consecutive arrests with minimum duration of $\Delta t = 125$ ms. Fig. \ref{figs:2}a gives an example the definition of $S$. Fig. \ref{figs:6}a shows the distribution of avalanche sizes for $w/d=2.33$ with 100 trials and 84 particles each. In plotting Fig. \ref{figs:6}a, as in Fig. \ref{figs:3}b, we use a logarithmic-binning method to provide better visualization of the tail of the distribution \cite{newman2005power}.  The 8 bins follow the rule: $S_{n+1}=S_n+1.7^n$, $0\leq n \leq 8$ and $S_0 = 0$, spanning the range $0\leq S \lesssim 98$. We plot data for $0 \leq \phi_r \leq 0.67$ only, because the number of instances for other $\phi_r$ values drops below 100 and does not provide enough statistics. As seen in Fig. \ref{figs:6}a, all distributions decay exponentially except for the first data point at $S=1$ for the three traces $\phi_r=0$, $0.17$, and $0.33$, where the distribution initially grows. This growth in $N(S)$ may be due to the free flow of the particles.  The average decay slope is $\alpha=0.07$ between all traces.  Interestingly, the distribution of avalanche sizes in a rigid granular discharge from a rectangular silo is shown to include two different regimes: a small region with growing numbers for small avalanche sizes until reaching a mode and then an exponential decay for larger avalanche sizes \cite{zuriguel2005jamming}. There, avalanches are forced to happen using external air puffs which is different from our spontaneous avalanches. The duration of avalanches $\tau$ scales linearly with the avalanche size $S$ with different slopes based on the particle sizes, as shown in Fig. S4 in the supplementary information.


The complementary cumulative distribution function (CCDF) of the avalanche sizes is calculated from the data in Fig. \ref{figs:6}a, by the following:

\begin{equation}
    P(S_0) =  \int_{S_0}^{\infty} N(S^\prime) \,dS^\prime / P_0
    \label{eq:3}
\end{equation}

\noindent where $P_0 = \int_{0}^{\infty} N(S^\prime) \,dS^\prime$. The resulting distributions are plotted in Fig. \ref{figs:6}b for the data shown in Fig. 5a. The inset of Fig. \ref{figs:6}b also displays the magnitude of the exponent $\beta$ for each trace. It is notable that the exponents of the CCDF are quite close to the average exponent in Fig. \ref{figs:6}a, meaning $\beta=\alpha$, which is expected for an exponential distribution. A smaller exponent indicates a smaller population of data at smaller $S$ values. The exponent generally decreases in magnitude with $\phi_r$, meaning that smaller avalanches are more likely for lower $\phi_r$.  That is, while more soft particles in the mixture lead to a greater number of arrests, as seen in Fig. \ref{figs:4}, these arrests end with avalanches of smaller sizes, and fewer particles outflowing after the arrest.


\vspace{1.5em}

\subsection{F\lowercase{irst-falling particle}}
\vspace{1em}


Increasing the fraction of soft particles in a mixture increases the degree of intermittency of the flow of the mixture through a hopper. This is quantified by the increase in both the frequency of long waiting times between particle exits and the number of temporary clogs that lead to avalanches. However, the underlying mechanism causing increased intermittency and avalanches remains unexplained. Avalanches in a hopper occur from the spontaneous breakage of the blocking arch. Here, we attempt to identify the cause behind the arch disruption and pinpoint the initiation point from which the falling wave propagates within the mixture.  Intuition may suggest that soft particles are similarly responsible for arch breakage.

To address this question, we investigate each instance in which a temporary arch collapses to allow flow.  We measure and compare the number of times a soft or rigid particle falls first from a temporary arch. To identify the ``first falling'' particle, for simplicity, we measure the time at which each particle passes below the hopper exit level.  An alternative method could involve measuring which particle moves first.  However, this would require setting a threshold to define ``zero motion,'' and is likely to introduce noise into the measurement.  By simultaneous inspection of the videos alongside the analysis, identifying the first particle falling below the exit level provides a reasonable designation of the particle that leaves the arch first.  There are some avalanche events in which the particle that leaves the arch first does not exit first, either due to local rearrangements of the arch particles or due to impact with the side wall.  These instances occur in approximately $\sim5$\% of all avalanche events.

Here we note that temporary arches are defined as $\Delta t > \Delta t_c$, where $\Delta t$ indicates a pause in particle exits, as shown in Fig. \ref{figs:3}a.  In short-lived temporary arches, with $\Delta t$ shorter than approximately 125 ms, particles in or near the arch may still move, albeit without causing particles to exit the hopper.  Once temporary arches persist longer than approximately $\Delta t \sim 125$ ms, particle motion in the arch becomes rare.  The supplementary videos provide examples of local particle motion in very short-lived arches.

\begin{figure}[h]
\centering
  \includegraphics[width=0.6\columnwidth]{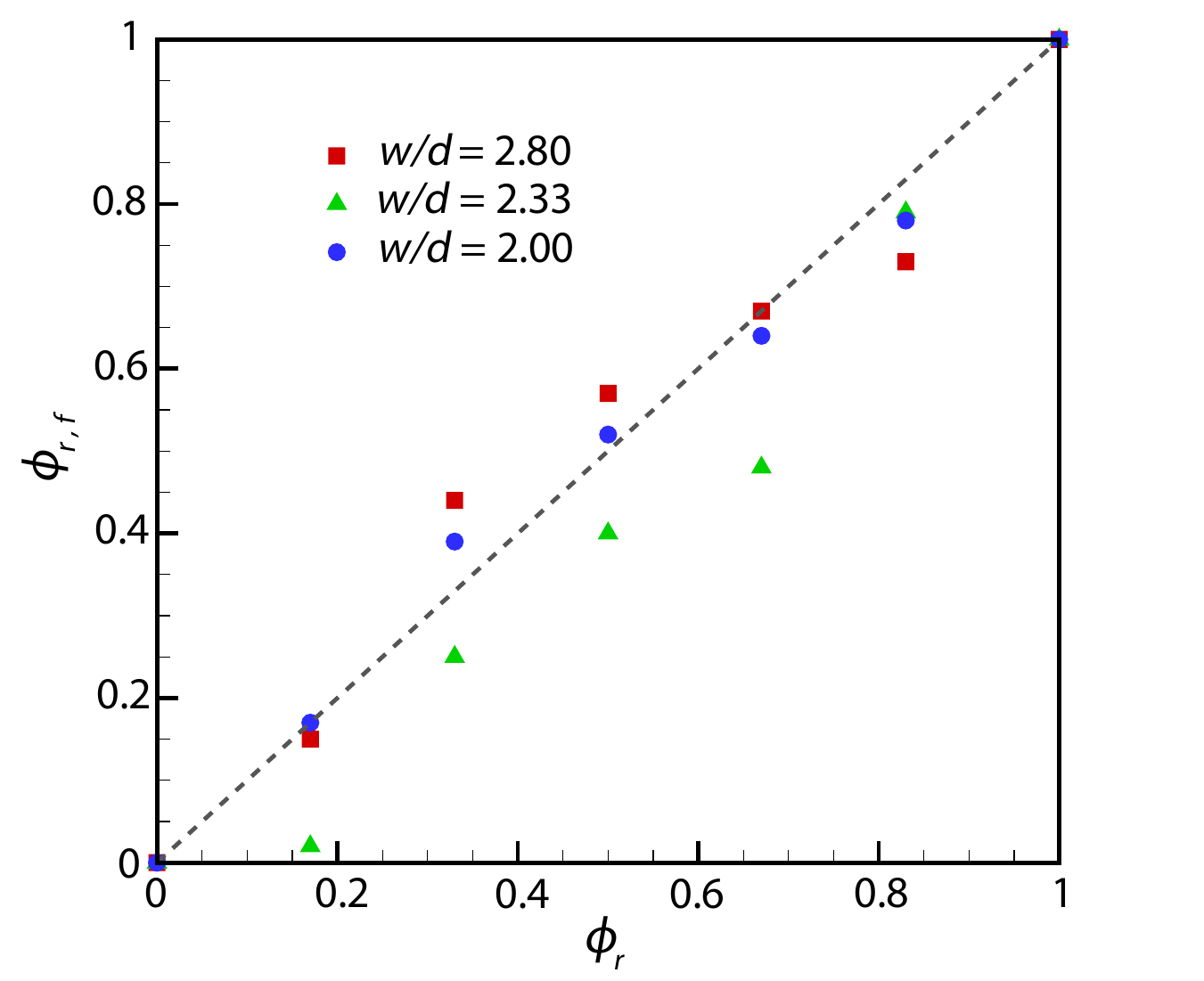}
  \caption{Fraction of times a rigid particle exits first after a temporary clog of time length $\Delta t \geq \Delta t_c$ versus the rigid fraction of the sample. The dashed line is only a 1:1 proportion for comparison.}
  \label{figs:7}

\end{figure}

To quantify the identity of first-fallers, we count the number of times a rigid particle falls first from a temporary clog of $\Delta t \geq \Delta t_c$ and normalize by the total number of temporary clogs within all trials for each $\phi_r$. Fig. \ref{figs:7} shows the percentage of first falling particles that are rigid, $\phi_{r,f}$ as a function of the rigid fraction of the entire system. 
Each data trace represents a different particle size. For each particle size, $\phi_{r,f}$ increases linearly with $\phi_r$ with a slope of nearly 1. The data for both $w/d=2.80$, in red, and 2.00, in blue, lie closest to the line where $\phi_{r,f}=\phi_r$. Interestingly, the majority of measurements obtained at $w/d=2.80$ and 2.00 fall above the 1:1 line, suggesting that rigid particles fall first. This is counter to our intuition that a soft particle would fall first. Only the data obtained at the intermediate particle size, $w/d=2.33$, in green, falls below the 1:1 line.  


To confirm that the automated image analysis protocol correctly labels the identity of the first falling particle, we manually check a selection of the data.  Each of the 21 data points in Fig. \ref{figs:7} represents up to several hundred ``first falling'' particle events.  We choose four of the 21 cases, two of which lie on the line where  $\phi_{r,f}=\phi_r$ and two of which lie below it.  Within these, we manually check 10\% of the data to confirm the first falling particle identity.  The data sets near $\phi_{r,f}=\phi_r$ are accurately quantified by the automated image analysis.  However, in the measurements of $w/d=2.33$, both when $\phi_r=0.17$ and 0.67, the manual check reveals an undercounting of rigid first falling particles by 18 and 6 \%, respectively.  In no case do we find that the automated analysis overestimates the fraction of soft particles falling first.  In all, there is no strong preference for the identity of the first falling particle to be, on average, significantly different than the composition of the entire system. Here we note that the identity of the particles relates both to their softness and also to their friction coefficients: the soft particles also experience lower friction \cite{alborzi2022soft}.


The results shown in Fig. \ref{figs:7} run somewhat counter to an intuition one may have that softer \added{or less frictional} particles would fall first and initiate an avalanche. Particle softness may not strongly determine which particle moves out of the arch first.  However, while Fig. \ref{figs:7} indicates the identity of the first falling particle, it does not show the nature of the friction between the first-faller and its neighbors.  It is still possible that contacts with neighbors may be more important than the identity of the first-faller itself.  To investigate this, we identify the neighboring particles of all first falling particles that are not next to the wall of the hopper.  We then categorize the two neighboring contacts of each first-faller as rigid-rigid, rigid-soft, and soft-soft, and present the results in Fig. S5. Interestingly, even the identity of the neighbors of the first falling particle scales with $\phi_r$, regardless of $w/d$.

\begin{figure}[h]
\centering

  \includegraphics[width=0.49\columnwidth]{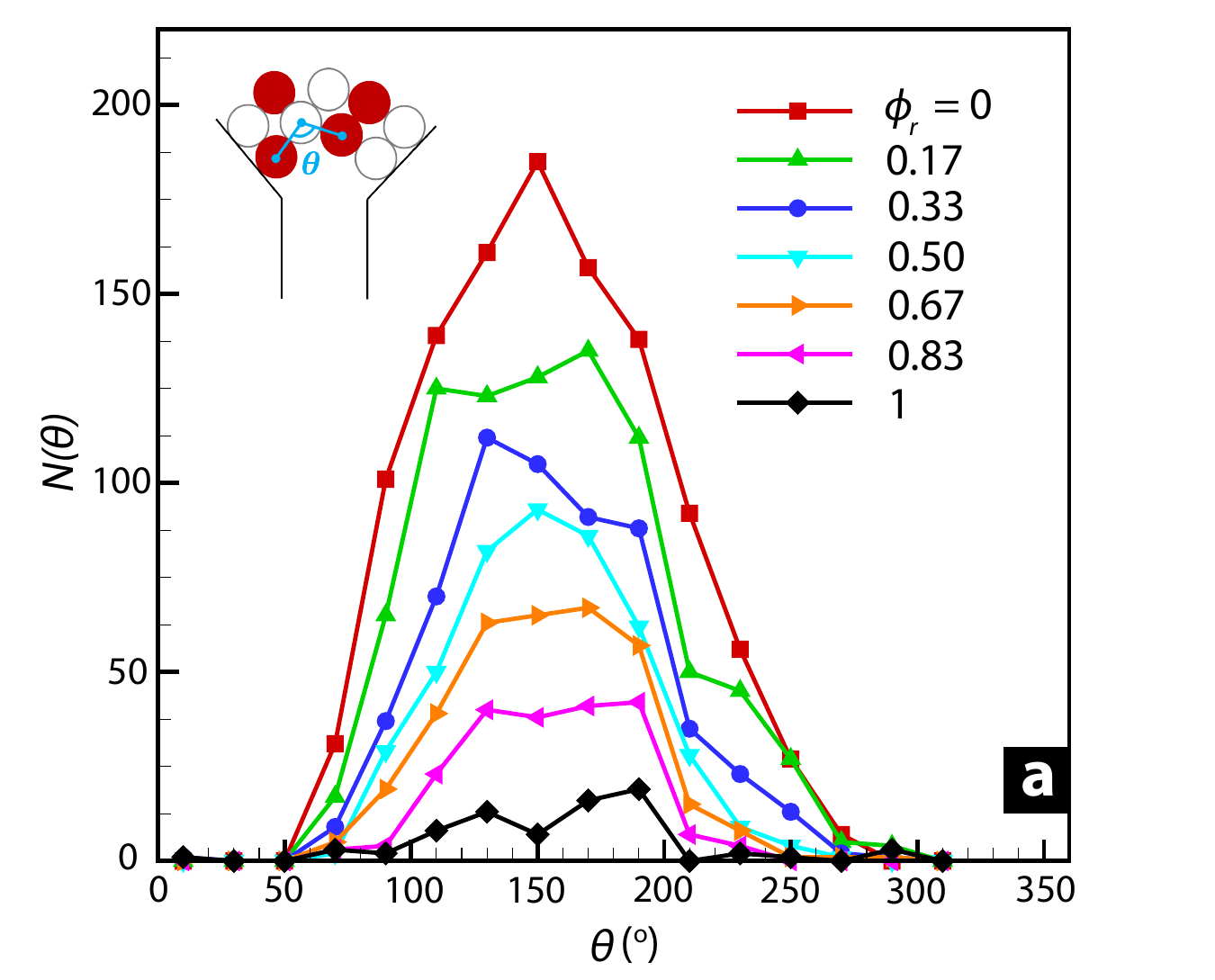}  
  \includegraphics[width=0.49\columnwidth]{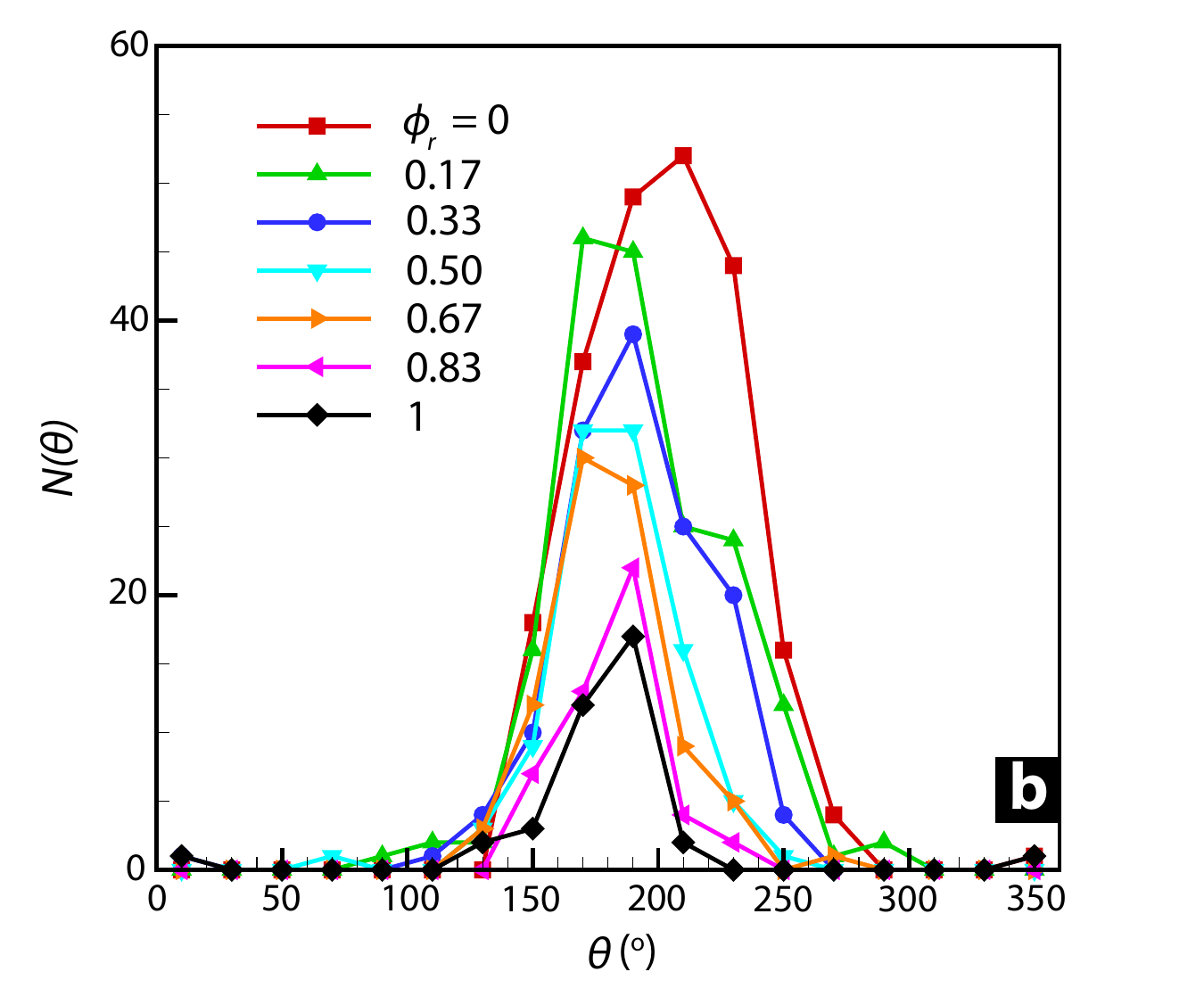}
  \caption{Distribution of angles between the particles in temporary arches of time length $\Delta t \geq \Delta t_c$ centering at (a) every arch particle and (b) the first-falling particle only, for different rigid fractions at $w/d=2.33$. Lines are to guide the eye only.}
  \label{figs:8}
\end{figure}

While the identity of the first-faller and its neighbors is mainly determined by system composition, the geometric shape of the arch reveals interesting features of arch instability. Arches are sometimes referred to as ``clogging microstates,'' and their geometry, including the position of any fixed particles, is a key determinant of clogging \cite{alborzi2023mixing,hanlan2024cornerstones}. Before discussing the meta-stable arches in our system, we first briefly discuss the geometry of permanent arches formed in granular systems of similar composition. In an earlier work, we show that the distribution of bond angles, $\theta$, between all neighboring particles in a permanent arch directly relates to the probability of clogging \cite{alborzi2023mixing}. The definition of the bond angle $\theta$ is shown in Fig. \ref{figs:8}a.  The distribution of angles both broadens and skews to larger angles as $\phi_r$ increases. Simultaneously the probability of clogging also increases.  Further, we observe two major peaks in the angle distribution in permanent arches: a peak at $\theta\simeq125^{\circ}$, indicating concave-shaped arches, occurs throughout the entire range of $\phi_r$ from 0 to 1.  A second peak at $\theta \simeq185^{\circ}$, indicating convex-shaped arches, appears as $\phi_r$ increases beyond $\sim0.6$, and corresponds to occurrences of three adjacent rigid particles in an arch.

Given the importance of arch geometry in permanent clogs, we now investigate the distribution of angles in temporary arches. To measure the angles between particles in a temporary arch, we choose the image captured seven frames, or $\sim58$ ms, after the \deleted{beginning of the temporary flow stoppage}\added{last particle exits in a flow arrest}. \added{Video2 in the supplemental materials shows this delay clearly. }At this time point, \deleted{the motion in the system is minimal}\added{the temporary arch has formed already}. Fig. \ref{figs:8}a shows the distribution of the angles in all temporary arches of duration $\Delta t\geq \Delta t_c$ for different $\phi_r$ values at $w/d=2.33$. The data shown corresponds to 550, 396, 313, 227, 170, 96, and 68 temporary arches observed during the 100 trials of each value of $\phi_r$ from 0 to 1, respectively. As seen in Fig. \ref{figs:8}a, all distributions range between $50^{\circ}\lesssim \theta \lesssim 300^{\circ}$ for every $\phi_r$.  The distributions are all roughly symmetric around $\theta=150^{\circ}$, with the magnitude of the peak decreasing from $\phi_r=0$ to 1. Thus, the curvature in unstable arches remains roughly constant regardless of the identity of particles.  In contrast, the distribution of angles in permanent arches exhibits a symmetric distribution of angles for $\phi_r \leq 0.20$ only, with a peak at $\theta\simeq125^{\circ}$ \cite{alborzi2023mixing}.  





In investigating the distribution of particle angles in the arch, the comparison between temporary and permanent arches does match our intuition. Angles above $180^{\circ}$, indicating a locally convex shape, do not appear in permanent arches except when $\phi_r > 0.5$.   Angles above $180^{\circ}$ do appear in temporary arches for all values of $\phi_r$.  This implies, intuitively, that particle contacts in temporary arches are more unstable and thus prone to disruption.  This meta-stability eventually leads to the escape of a particle and initiation of an avalanche. 


While Fig. \ref{figs:8}a shows the distribution of particle angles for all particles in a temporary arch, we also investigate the angle formed by the first falling particle and its neighbors.  We plot the distribution of angles centered at the first-falling particle in Fig. \ref{figs:8}b for the same particle size $w/d=2.33$ shown in Fig. \ref{figs:8}a. In Fig. \ref{figs:8}b, however, we exclude first falling particles which are next to the wall.  Although the population of data in Fig. 7b is smaller, the distributions show a clear shift to larger angles. In Fig. \ref{figs:8}a, almost all distributions peak at $\theta \simeq 150^\circ$, while in Fig. \ref{figs:8}b the peaks occur between $\theta\simeq 170^\circ$ to $210^\circ$ for different $\phi_r$. This implies that, when investigating the ensemble of all temporary arches, the bond angle of the first-falling particle with its neighbors takes on a larger value than other angles in the arch. 



The comparison between the distributions in Fig. \ref{figs:8}a and Fig. \ref{figs:8}b suggests that, within a given temporary arch, the particle experiencing the largest bond angle between its neighbors is the particle that will fall first.  To test this prediction, we investigate all temporary arches of four or more particles.  With four or more particles in an arch, there are at least two bond angles to compare.  We also limit our investigation to cases in which the first falling particle is not on the sidewall.  In this way $\theta$ is the bond angle between three particles, as shown in Fig. \ref{figs:8}a. We refer to this subset of first falling particles as \textit{middle particles} and to the corresponding arches as \textit{target arches}. 


Table \ref{table:2} shows the number of target arches in all 21 conditions and the percentage of arches in which the particle with the maximum bond angle falls first. As seen from the \%True column, the assumption is proved correct between 68.09\% and 96.88\% in all conditions, with a weighted average of 85.35\%. These calculations, however, are subject to some error due to two main reasons: 1. In some occasions, the first falling particle impacts one of the inclined walls and thus exits after another particle.   2. Sometimes, the particle which falls first from the arch is lost from tracking near the exit and therefore a different, neighboring particle is mistakenly identified by the code as the first to exit. By manual inspection, we correct the mistakes due to these two reasons and list the new results as ``modified'' in the table.  In some cases, the number of target events changes because a middle particle was misidentified as the first-faller, rather than a side particle. Comparing the original number of target arches with the modified number shows that only three arches out of 274 are removed for the measurements at $w/d=2.8$. Twelve arches out of 806 are removed for the measurements at $w/d=2.33$. Only one arch is removed out of the 360 measurements at $w/d=2.00$.  After this manual adjustment is made, the weighted average rises to 92.84\%. That is, in nearly 93\% of temporary arches, the particle that falls first is the particle with the largest bond angle with its neighbors.  In the remaining $\sim7$\% of temporary arches, which the first-faller does not have the largest $\theta$.  This can be explained largely by the observation that temporary flow stoppages, especially short-lived ones, may be accompanied by local motion of individual particles in or near the arch.  More specifically, we identify two reasons: 1. We measure $\theta$ at a time point seven frames beyond the flow blockage.  While this is a reasonable choice, the system is occasionally still rearranging to find its meta-stable form. Thus, the angles between the arch particles in that particular frame might still change slightly over the next few frames. 2. In some cases, the first falling particle has $\theta$ only slightly lower than the maximum angle, by roughly $< 10^{\circ}$.  This may again be due to rearrangements throughout the system. 

\begin{table}[!h]
\centering
\tiny
\caption{Number of target arches for predicting the first-falling particle and the percentage of instances where the maximum angle aligns with the first-falling particle at each particle size and rigid fraction.}
\label{table:2}
\resizebox{0.8\columnwidth}{!}{
\begin{tabular}{c c c c c c}
\hline \\ [-2pt]
 \multirow{3}{*}{$w/d$} & \multirow{3}{*}{$\phi_r$} & \multirow{3}{*}{\shortstack[c]{Target \\ arches}} & \multirow{3}{*}{\%True} & \multirow{3}{*}{\shortstack[c]{Target \\ arches \\ (modified)}} & \multirow{3}{*}{\shortstack[c]{\%True \\ (modified)}} \\ \\  &&&&& \\[5pt] \hline \\ [-3pt]

  2.80 & 0 & 31 & 90.32 & 31 & 90.32 \\ & 0.17 & 18 & 83.33 & 18 & 100 \\ & 0.33 & 32 & 87.50 & 31 & 93.50 \\ & 0.50 & 53 & 84.91 & 53 & 88.68 \\ & 0.67 & 59 & 79.66 & 58  & 91.38 \\ & 0.83 & 47 & 68.09 & 47 & 87.23 \\ & 1 & 34 & 79.41 & 33 & 84.85 \\ 2.33 & 0 & 221 & 90.95 & 221 & 93.67 \\ & 0.17 & 176 & 84.09 & 173 & 95.95 \\ & 0.33 & 136 & 86.76 & 132 &  95.45 \\ & 0.50 & 99 & 85.86 & 97 & 93.81 \\ & 0.67 & 88 & 76.14 & 88 & 92.50 \\ & 0.83 & 48 & 83.33 & 47 & 93.62 \\ & 1 & 38 & 76.32 & 36 & 86.11 \\ 2.00 & 0 & 32 & 96.88 & 32 & 96.88 \\ & 0.17 & 59 & 81.36 & 59 & 86.44 \\ & 0.33 & 34 & 94.12 & 33 & 100 \\ & 0.50 & 59 & 88.14 & 59 & 93.22 \\ & 0.67 & 69 & 86.96 & 69 & 91.30 \\ & 0.83 & 45 & 93.33 & 45 & 95.55 \\ & 1 & 62 & 87.10 & 62 & 90.32

\end{tabular}}

\end{table}

We find these observations to be both interesting and surprising.  The presence of soft particles in granular mixture increases both the frequency of intermittent flow events and the duration of flow stoppage.  We might expect that some combination of more softness and less friction would strongly contribute to restarting flow after a temporary stoppage. However, the particle that falls first after temporary pauses in flow is not necessarily more likely to be a soft particle. That is, particle identity does not necessarily determine the least stable position in the blocking arch.  In fact, even having contact with a soft particle neighbor does not mean that any particle is more likely to fall first.  Instead, we find geometry to be the strongest determinant of instability in the system.  Bond angles in arches measured during temporary pauses in flow exceed those found in permanent clogs.  Not only does temporary arch geometry differ from that in permanent arches, but also it determines the onset of instability.  It is the largest bond angle that is the most unstable, regardless of the identity of the particle at the center of the bond or the identity of its neighbors. The particle with the largest bond angle is the most likely to fall first, thus restarting flow.





\section{Conclusions}

Apart from preventing permanent clogging, introducing soft particles into rigid granular systems adds further interesting phenomena. Rigid particles display a fast, non-intermittent flow behavior unless the system clogs permanently. In contrast, the flow of soft particles shows variable time lapses between the exiting particles $\Delta t$, with larger average values than those in rigid systems. However, the addition of soft particles also reduces the probability of permanent clogs in the system.  As such, mixtures containing soft particles exhibit faster average discharge rates over the duration of a flow test.

Measurements of avalanche sizes and time intervals between the particle exits reveal some unexpected results.  While flow pauses occur regardless of the mixing fraction, they occur more frequently and persist for longer durations when soft particles predominate the mixture. These pauses may happen due to the presence of adhesion in the particle contacts. We measure lower adhesion and friction in soft-soft contacts compared to soft-rigid and rigid-rigid contacts, which may enable more fragile arches at low $\phi_r$. Avalanche sizes are distributed exponentially, and the distribution of time lapses between falling particles is distributed on a power law, similar to behavior seen in other granular systems. Interestingly, some of the longest time lapses between particle exits occur when the mixing fraction is intermediate between soft and rigid.




Given that avalanches and intermittents events happen more frequently with soft particles in mixture, intuition might suggest that soft particles have the hardest time staying in place in an arch, and therefore fall first to initiate an avalanche.  Therefore we investigate different mechanisms behind the initiation of avalanches and focus on the identity of the first falling particle. However, we find the probability that an avalanches is initiated by a soft particle is not significantly different than the mixing ratio of soft particles in the entire system.  Further, quantifying neighbor contacts of the first falling particle also does not indicate any predictive power for avalanches.



Even though the identity of the first falling particle is not strongly correlated with avalanche occurrences, we do find a possible explanation by investigating the geometry of the temporary arches.  We analyze the distribution of angles $\theta$ between the arch particles for the cases where $\Delta t>\Delta t_c$. The distributions of all bond angles in a temporary arch is nearly normally distributed around $150^{\circ}$, regardless of $\phi_r$.  When we focus on the bond angle made by the first falling particle and its neighbors, the distributions show an average value of $\theta=170^{\circ}$ or larger. This distinction implies that the first-falling particle is the one that takes the largest angle among all particles in the temporary arch.  This geometric instability is what is strongly correlated with, and likely leading to, the occurrence of avalanches. We evaluate this claim by comparing the bond angle of the first falling particle to all other bond angles within each temporary arch.  Indeed, considering all data investigated, the first falling particle does have the largest bond angle within a temporary arch in $\sim93$\% of all avalanche events.


These experimental observations suggest a number of open questions. In particular, the observation that the first falling particle identity is not strongly correlated with avalanche occurrence could be investigated in more detail.  The two species in this experiment behave similarly in terms of which particle falls first to cause an avalanche.  However, there may be other types of particles that would more clearly distinguish the behavior of ``soft'' from ``rigid'' particles.  
These questions might be more easily approached through simulations of granular hopper flows than through experiments. That is, investigating a range of different particle moduli in mixed systems could reveal the dependence of avalanche onset on particle softness as a continuously controlled parameter.  Similarly, investigating particles with controllable sliding friction factors could reveal situations in which neighbor contacts become important in stabilizing avalanches.  Despite these open questions, the observation that arch geometry is important in determining avalanche onset may have implications in the study, and perhaps even prediction, of avalanches in a wide variety of granular flows.

\section{CR\lowercase{edi}T \lowercase{authorship contribution statement}}

\textbf{Saeed Alborzi:} Conceptualization, Methodology, Experiments, Data Analysis, Writing -- original draft \& editing. \textbf{Sara M. Hashmi:} Conceptualization, Methodology, Project administration, Writing -- reviewing \& editing, Funding acquisition, Supervision. 

\section*{A\lowercase{cknowledgments}}

The authors wish to thank Eric Weeks and Katherine Jensen for helpful discussions.

\section*{D\lowercase{eclaration of Competing Interest}}

The authors have no conflicts of interest to declare.

\vspace{1em}


%

\end{document}


\title{Supplementary Information}

\author{Saeed Alborzi}
\affiliation {Department of Mechanical and Industrial Engineering, Northeastern University, Boston, MA 02115, USA.}
\author{Sara M. Hashmi}
\altaffiliation{s.hashmi@northeastern.edu}
\affiliation {Department of Mechanical and Industrial Engineering, Northeastern University, Boston, MA 02115, USA.}
 \affiliation{Department of Chemical Engineering, Northeastern University, Boston, MA 02115, USA.}
 \affiliation{Department of Chemistry and Chemical Biology, Northeastern University, Boston,
MA 02115, USA.} 

\maketitle




\date{\today}


\section{A\lowercase{verage particle velocity at exit}, $\bar{v}_{out}$}

In a rotating hopper, particles' outflow velocity undergoes different stages during the discharge process. Fig. S1 shows the variation of average particle velocity at the exit, $V_{out,m}$, versus the number of exited particles at each step for three various $\phi_r$'s at $w/d=2.80$. The data are averaged at each $N_{exited}$ value between the all trials. The initial number of particles $N_{exited}<10$ exit while hopper is still rotating before it reaches the upright position, thereby moving at a higher velocity than when the hopper is stationary. After stopping at upright position, $V_{out,m}$ reaches to a steady state. The discharge process continues until the fill height reduces below a minimum $N_{exited} \gtrsim 80-90$ when the above weight and base pressure drops significantly. Therefore, lower driving pressure results in lower outflow velocities too. This behavior is seen in all $\phi_r$ values shown. Additionally, as $\phi_r$ increases, the velocity data is truncated at lower values of $N_{exited}$ due to the effect of permanent clogs.  For all systems studied, at least 80\% of particles exit the hopper in a steady flow regime.  As $\phi_r$ decreases to 0, the steady flow regime increases slightly.  

\begin{figure}[h]
\centering
  \includegraphics[height=8cm]{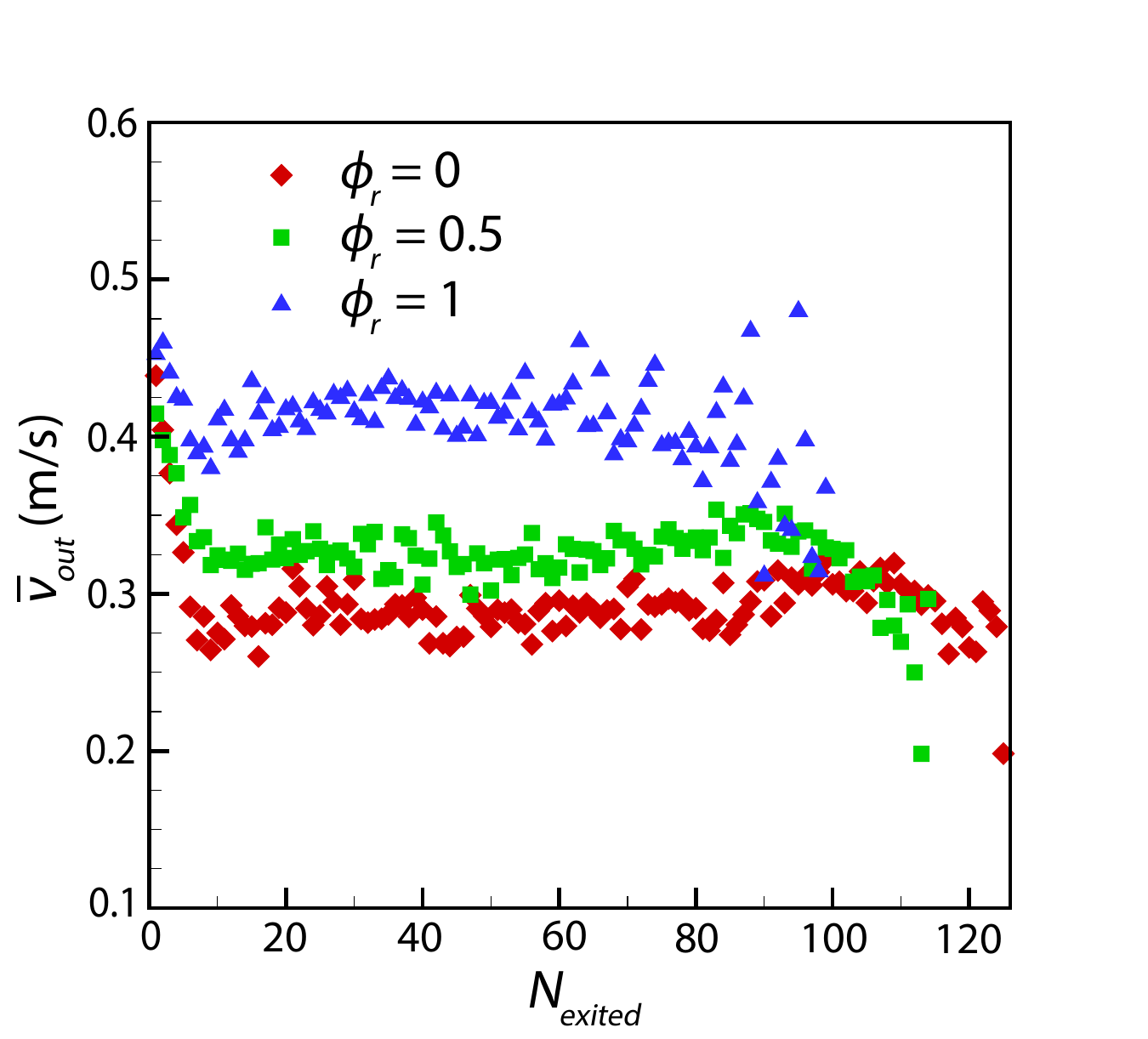}
  \caption{Variation of average particle velocity at hopper exit vs. number of exited particles for $w/d=2.80$}
  \label{figs:SI1}
\end{figure}

\newpage

\section{A\lowercase{verage number of drained particles}}

To compare the overall number of particles that exit prior to a permanent clog in a single trial on average, we plot the average difference between $N_{rem}(t=0)$ and $N_{rem}(t\rightarrow\infty)$ over all trials against $\phi_r$ for the data shown in Fig. 2c. As depicted in Fig. S2, $\overline{\Delta N}_{rem}$ declines as the rigid fraction increases, except for $\phi_r=0.83$ at which the slight rise is within the error bar sizes. As the portion of rigid particles increases in the mixture, the probability of permanent clogging rises and thus the amount of exited particles is limited. Although the velocity of single particles $v_{out}$ is larger for rigid particles from the results shown in Fig. 2b, the higher clogging probability in pure rigid systems leads to lower overall discharge amounts. This phenomenon is commonly referred to as \textit{faster-is-slower.}

\begin{figure}[h]
    \centering
        \centering
        \includegraphics[height=8cm]{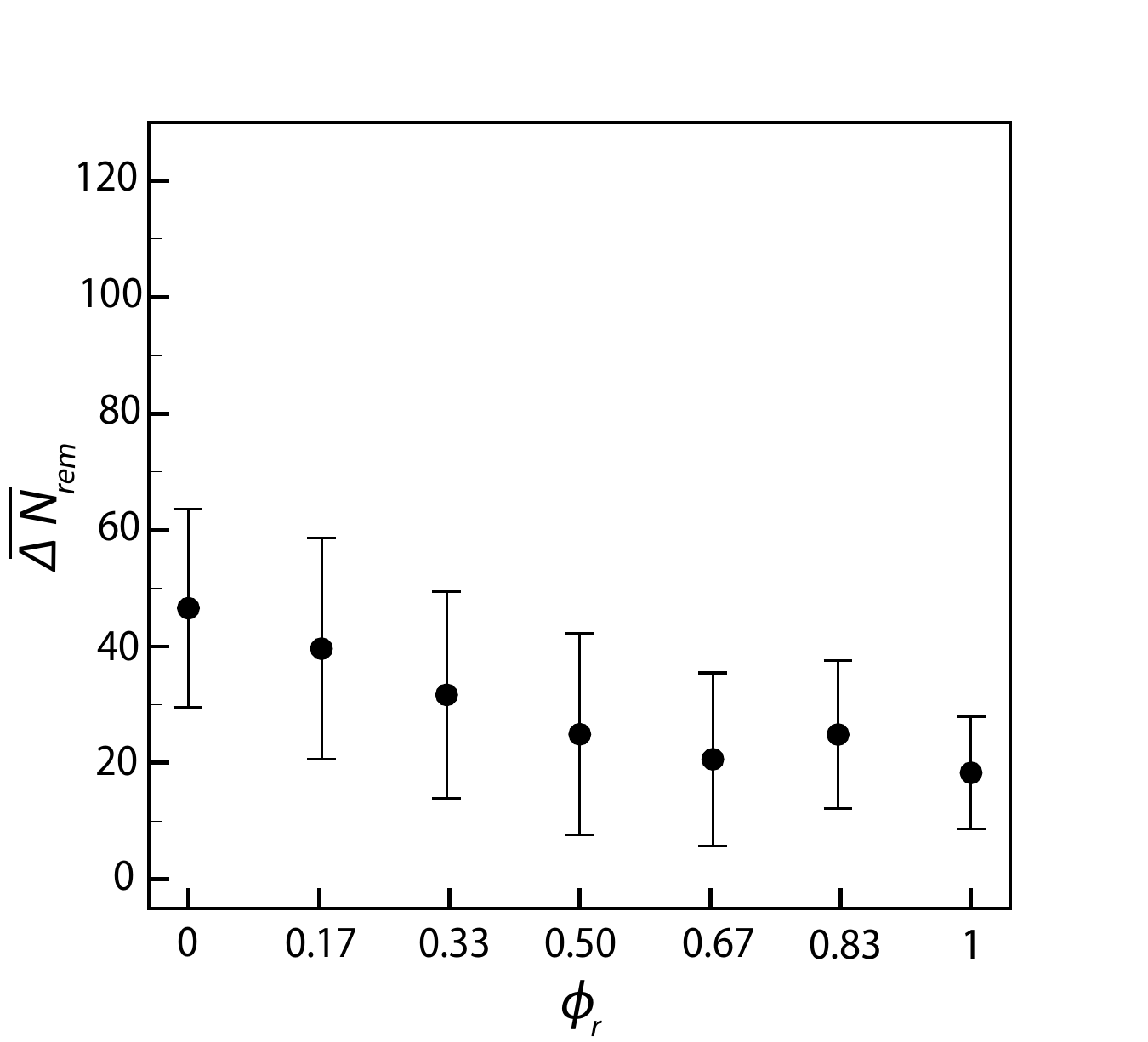}
    \caption{Average number of exited particles $\overline{\Delta N}_{rem}=N_{rem}(t=0)-N_{rem}(t\rightarrow\infty)$ before a permanent clog occurs for different rigid fractions at $w/d=2.00$. Each point represents the average over 150 trials and the error bars show the standard deviation.}
    \label{fig:panel}
\end{figure}

\newpage

\section{T\lowercase{ime lapses $\Delta t$ scatter plot during the discharge}}


The dispersion of time lapses $\Delta t$ over the entire discharge duration is shown in Fig. S3, for $w/d=2.33$ and three rigid fractions $\phi_r=0$, $0.50$, and $1$. Each data point represents the event of a single particle exiting, for all 100 rotations. The cyan dashed line corresponds to the critical time lapse value $\Delta t=125$ ms, above which the instances are considered temporary clogs. The $x$-axis is normalized by the total number of particles $N_t=84$ for this particle size. Due to the presence of clogs at the end of individual runs, the total number of data points shown decreases from $\sim8000$ at $\phi_r=0$ to $\sim2000$ at $\phi_r=1$.  At $\phi_r=0.5$ and $1$, the data truncates before $N_{exited}/N_t=1$, largely due to the presence of these clogs at the end of a run.  I a few runs at $\phi_r=0.5$ and $1$, tracking errors may cause the loss of a few particles that exit at high velocity.  However, we anticipate the number of particles lost in tracking is small compared to the number of clogs that remain in a clog. 

Fig. S3 shows an interesting effect due to mixing.  For rigid systems, in blue, the distribution is skewed toward the start of the experiment, where the flow is still transient. Larger time intervals between the earliest exiting particles likely correspond to particles exiting while the hopper is still rotating.  When $\phi_r=0$, in red, the longest intervals between particle exits occur near the end of the run.  This corresponds to the slightly slower flow in soft systems when less than $\sim10$ particles remain, as seen in Fig. S1.  However, when $\phi_r=0.5$, in green, with an equal number of soft and rigid particles in the mixture, the instances of particles exiting with $\Delta t>125$ ms occur throughout the entire discharge time.  Most of these intermittent events occur during the steady portion of the outflow seen in Fig. S1.  The longest time intervals between particle exits also occur during steady outflow. Some intervals are very long, up to 3s in duration.  Interestingly, no intervals between particle exits are longer than 0.8s for purely rigid systems.  Intervals between particle exist exceed 1s for purely soft systems only in the unsteady final outflow of the final $\sim10$ particles.

\begin{figure}[h]
    \centering
        \centering
        \includegraphics[height=8cm]{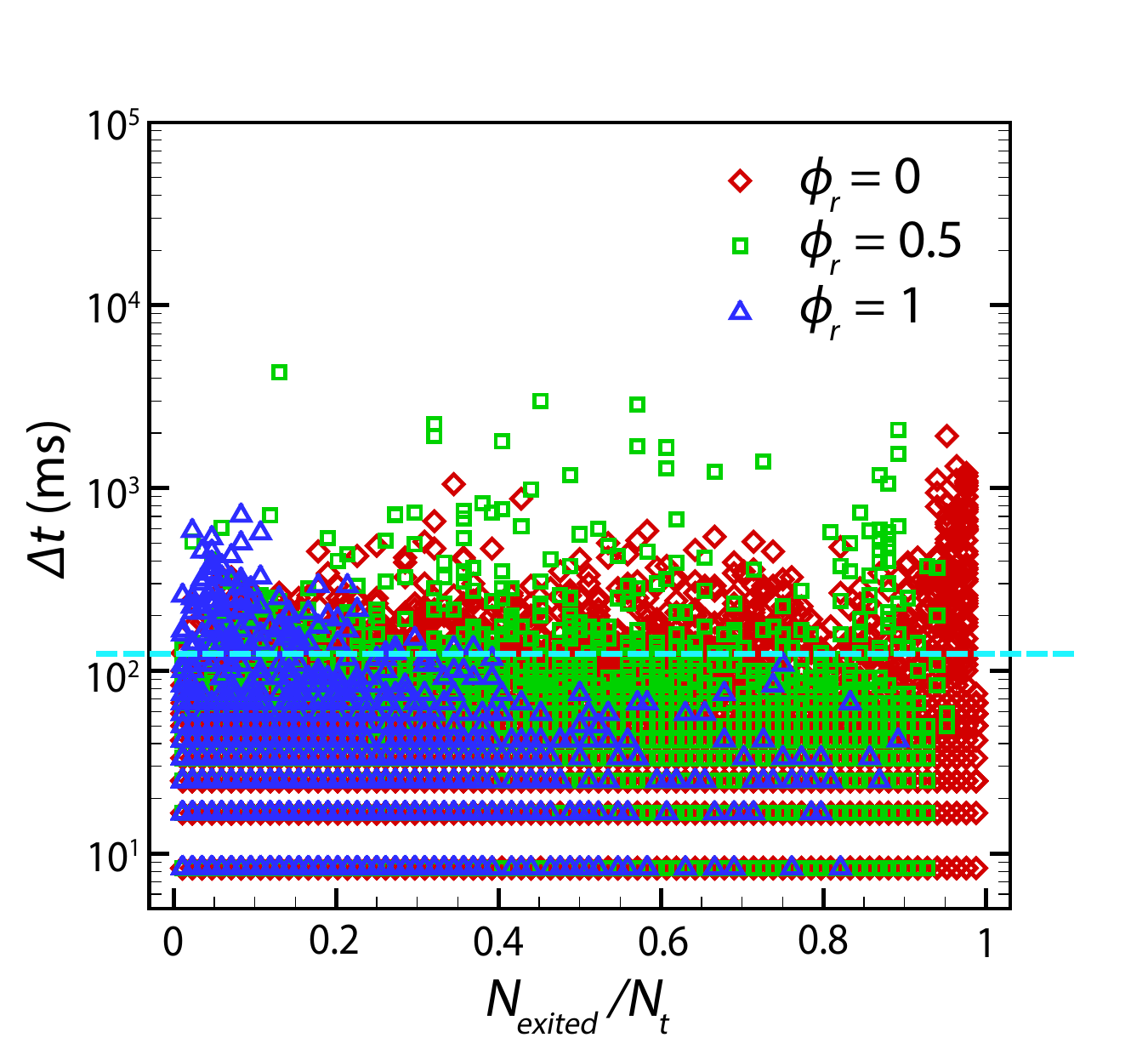}
    \caption{Time intervals $\Delta t$ vs normalized number of exited particles for $w/d=2.33$. $N_t$ is the total number of particles which 84 in this case. The cyan dashed line represents $\Delta t_c = 125$ ms.}
    \label{fig:panel}
\end{figure}

\newpage

\section{A\lowercase{valanche size (\MakeUppercase{S}) vs avalanche duration ($\tau$)}}


Fig. S4 shows the avalanche size $S$ as a function of the avalanche duration $\tau$ for the size ratio $w/d=2.33$, with a total number of 84 particles in the system. Avalanche duration $\tau$ refers to the length of time in which flow occurs between two plateaus in $N(t)$ as indicated in Fig. 2a.  Note that we define an avalanche as the number of particles discharged between any two temporary clogs of duration $\Delta t \geq \Delta t_c$, as indicated in Fig. 2a. The start or the end of each run can also be the start or the end of an avalanche. Here, we plot only three rigid fractions $\phi_r =0$, $0.5$, and $1$ to better visualize the distinctions between different datasets. 

The data generally shows a clear linear correlation between the avalanche size and duration, similar to previous studies \cite{hidalgo2018flow}.  Interestingly, similarly to Fig. S3, avalanches are more prominent in the mixture than in the homogeneous soft or rigid systems.  In rigid systems $\phi_r=1$, avalanches mostly occur over shorter durations, less than 1s.  These short duration avalanches often correspond to the instances between the start of a trial followed by a permanent clog. Only 4\% of data points lie beyond $\tau=1$s for $\phi_r=1$. As $\phi_r$ decreases, the distribution of data gets broader, and avalanches last for longer times. In an equal mixture, 21\% of avalanches last longer than 1s, and some as long as $\sim2.4$s. When $\phi_r=1$, fewer avalanches last longer than 1s, only 9\%.  However, the longest few avalanches last even longer than 2.4s. The linear slope of $S(\tau)$ decreases as $\phi_r$ decreases.  This is due to the decrease in average particle velocity, as shown in Fig. 2b and Fig. S1. The slopes of the fitted lines in Fig. S4 are approximately 0.025, 0.031, and 0.034 $ms^{-1}$ for $\phi_r=0$, $0.5$, and $1$, respectively.
\begin{figure}[h]
\centering
  \includegraphics[height=8cm]{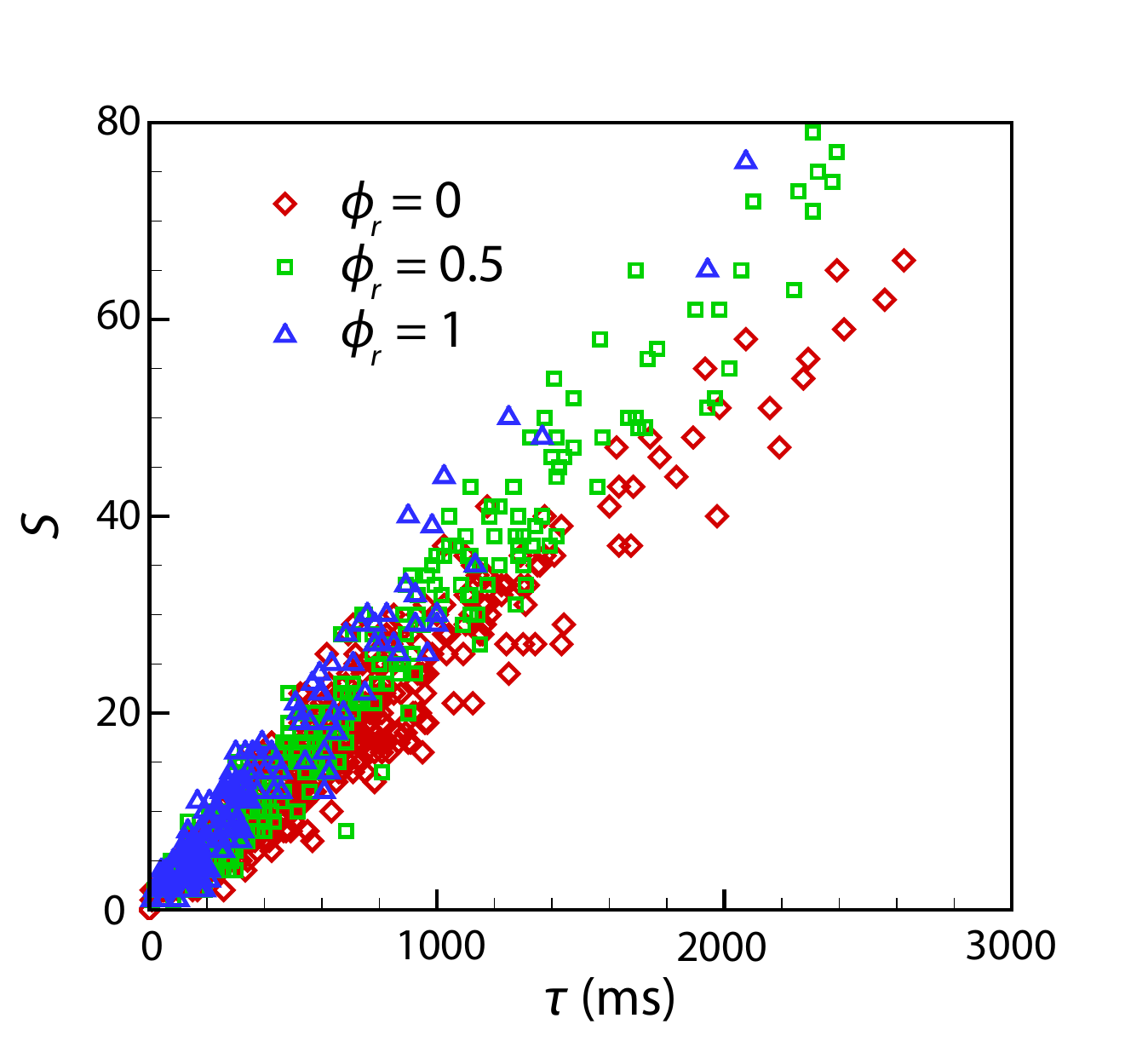}
  \caption{Avalanche size $S$ vs avalanche duration $\tau$ for $w/d=2.33$.}
  \label{figs:SI2}
\end{figure}

\newpage


\section{C\lowercase{ontact types of the first-falling particle with neighbors}}

In the cases that one of the middle particles falls first from a temporary arch, the contact types with its neighbors can be quantified and compared across different $\phi_r$ values. There are three possible contact types when considering the first falling particle and two neighboring particles: soft-soft (SS), rigid-rigid (RR), and rigid-soft (RS). To measure the probability of each type, we divide the number of events for each contact type by the total number of contacts at each $\phi_r$. Fig. S5 shows the probability of each contact type for every $\phi_r$ between 0-1 at $w/d=2.00$. The results show a proportional variation with respect to $\phi_r$.  That is, $P(RR)$ increases monotonically with increasing $\phi_r$ or $P(RS)$ peaks at $\phi_r=0.5$ and decreases on both sides. This behavior implies that the particle that falls first is not strongly determined by the type of its contact with its neighbors.

\begin{figure}[!htbp]
    \centering
    \begin{minipage}{.49\textwidth}
        \centering
        \includegraphics[width=\linewidth]{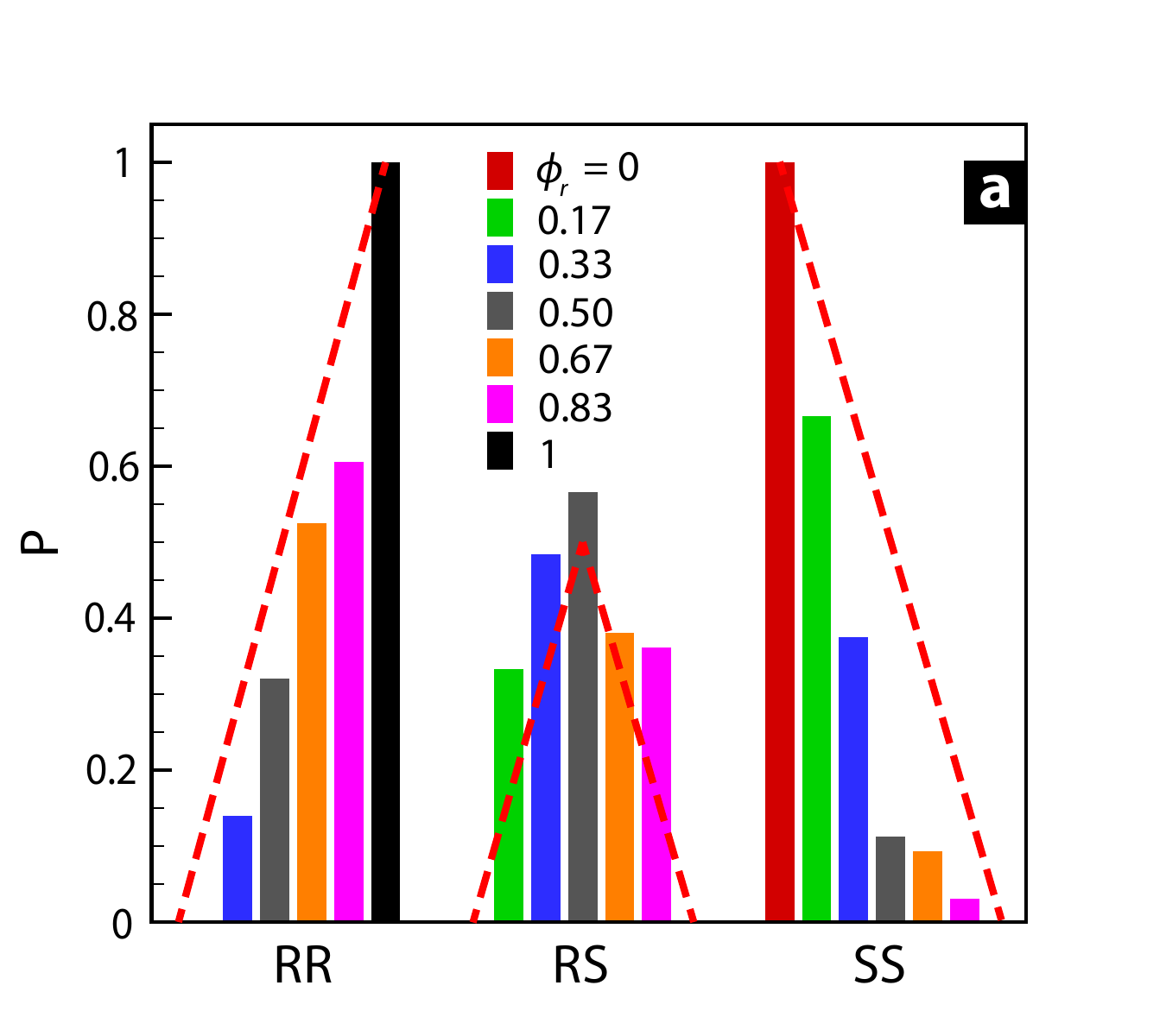}
    \end{minipage}%
    \begin{minipage}{.49\textwidth}
        \centering
        \includegraphics[width=\linewidth]{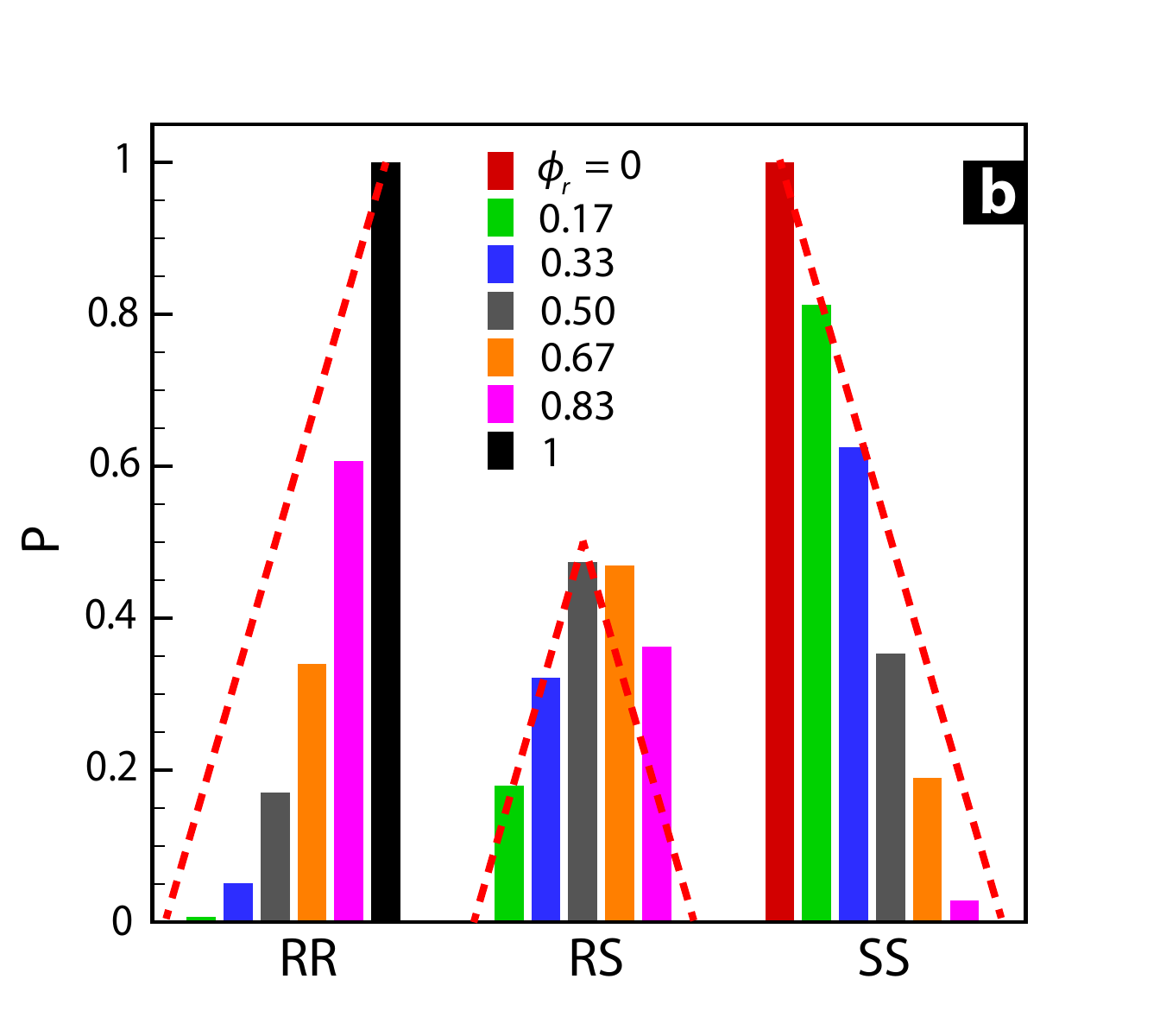}
    \end{minipage}
    \begin{minipage}{.49\textwidth}
        \centering
        \includegraphics[width=\linewidth]{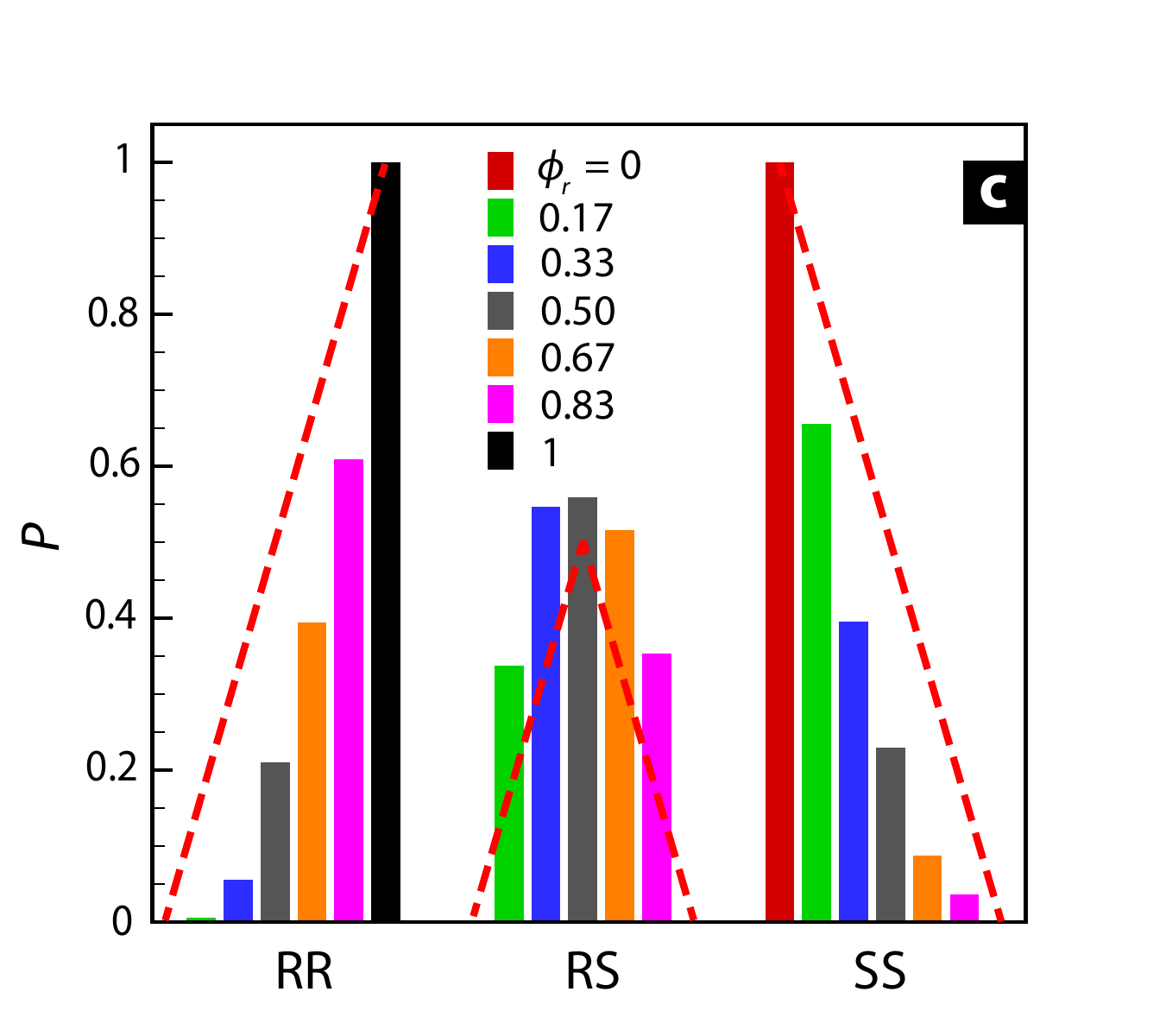}
    \end{minipage}
    \caption{Probability of different contact types around the first-falling particle for various rigid fractions at (a) $w/d=2.80$, (b) $w/d=2.33$, and (c) $w/d=2.00$. 'RR', 'RS', and 'SS' represent the rigid-rigid, rigid-soft, and soft-soft contacts, respectively. The red dashed lines in all figures show 1:1 ratio between the probabilities and rigid fraction.}
    \label{fig:panel}
\end{figure}

\newpage

\subsection{S\lowercase{upplementary Videos}}

The following video files can be found at: \url{https://github.com/smhashmi786/hashmilab_public/tree/main/Avalanche%20Videos}

\textbf{Video1.mp4:} This video is a combination illustrating the difference between four different durations of $\Delta t$, the time interval between exiting particles: 83.3, 125, 166.7, and 208.3 ms. A yellow box appears around the blocking arch for the duration of each time interval during which no particle exits. After viewing many videos of the entire range of $\Delta t$ values observed in all systems measured, we determine that time lapses $\Delta t<125$ ms do not constitute full arrest. Therefore we choose the critical time lapse value as $\Delta t_c = 125$ ms. That is, the video for $\Delta t = 83.3$ ms is not an instance of full arrest, nor is there an avalanche following $\Delta t = 83.3$ ms.  The other three videos do show instances of arrest followed by avalanches. The videos are timed so that the beginning of $\Delta t$ occurs at the same time in each: the yellow boxes appear simultaneously. In each video the playback frame rate is 30 fps, compared to the actual frame rate of 120 fps. The system shown corresponds to $w/d=2.80$ and $\phi_r=0.67$.

\textbf{Video2.mp4:} This video shows the moment occurring seven frames, or 58.3 ms, after the last particle exits and before the onset of an avalanche.  Both the moment at the beginning of the arrest and the moment seven frames afterward are shown in the video by momentary blue and yellow boxes, respectively. The video playback pauses on each of these frames, for emphasis.  The angles between the arch particles are measured in the frame with the yellow box. Also, the first-falling particle following the arrest is circled in red, and the text shows the computer-identified type of this particle. The playback frame rate is set to 6 fps for better visibility, compared to the actual frame rate of 120 fps. The system shown corresponds to $w/d=2.00$ and $\phi_r=0.50$.